\documentclass[twocolumn,showpacs,preprintnumbers,amsmath,amssymb, floatfix]{revtex4-1}

\usepackage{graphicx}
\usepackage{dcolumn}
\usepackage{bm}

\usepackage{physics}
\usepackage{subcaption}
\usepackage{xcolor}

\begin{document}

\title{Theoretical description of the ionization processes with a discrete basis set representation of the electronic continuum}

\author{Mateusz~S.~Szczygie\l}
\email{mateusz@tiger.chem.uw.edu.pl}
\author{Micha\l~Lesiuk}
\author{Robert~Moszy\'{n}ski}
\affiliation{
Chemistry Department, University of Warsaw, Pasteura 1, 02-093 Warsaw, Poland
}

\date{\today}

\begin{abstract}
In this work we present a new method of approximating the continuum wavefunctions with a~discrete basis set.
This method aims to be at least compatible with other well known methods of the electronic structure theory to describe processes in the electronic continuum.
As an example of the application we present the results of the calculations of the photoionization cross sections for the hydrogen atom and hydrogen molecule and for the helium atom.
We also obtained the photoelectron angular distribution for the hydrogen molecule.
The agreement of our results with the experimental data is very good in the energy range from the ionization threshold up to photon energies of about 60~eV.
\end{abstract}


\maketitle

\section{\label{sec:introduction} Introduction}

The process of photoionization has been the object of many experimental and theoretical study.
For the hydrogen atom the exact result was given by Stobbe \cite{stobbe1930}.
The simplest model describing the photoionization process consists of approximating the ejected electron with the plane wave and the corresponding results can be found in standard textbooks \cite{Schiff}.
For the hydrogen-like atoms this gives the correct results for the S-states \cite{Bethe}.
In the case of molecular photoionization cross section the discussion about using plane waves and Coulomb wavefunctions as an approximate continuum orbitals is given by Gozem \emph{et al.} \cite{gozem2015}.
Another standard method is the Born approximation \cite{Landau, Bethe} in which one approximates the total wave function, given as a solution of the Lippmann-Schwinger equation \cite{Weinberg, Lippmann1950}, to the first order.

The methods involving finite basis set expansions are the complex coordinates method \cite{rescigno1975}, with the notable application to the helium atom \cite{stark2011}, and the complex-valued basis functions \cite{russell1985, matsuzaki2017}.
A recent application of the latter to the calculation of the cross sections for the hydrogen and helium atoms, and the hydrogen molecule was published by Morita and Yabushita \cite{morita2008, morita2008_h2}.
These methods rely on the fact that the total photoabsorption cross section can be written as \cite{fano1968} 
$$
\sigma = \frac{4 \pi \omega}{c} \mathrm{Im} \: \alpha^- (\omega),
$$
where $\alpha^- (\omega)$  is the frequency dependent polarizability
$$
\alpha^- (\omega)  = \mel{\Psi_0}{d \frac{1}{H - E_0 - \omega - i0} d }{\Psi_0},
$$
where $d$ is the dipole moment operator, $\Psi_0$ is the ground state of the molecular Hamiltonian $H$ with the energy $E_0$.
The incident photon energy is denoted by $\omega$.
Based on the above methods the Stieltjes-Chebycheff imaging can be used \cite{langhoff1974, cacelli1988} where one obtains  continuum oscillatory strengths from its discrete representation in a large basis set.
The advantage of these methods is that one avoids using the continuum orbitals explicitly.

More advanced studies utilize various techniques of the many-body perturbation theory \cite{kelly1987}.
The R-matrix method of Burke \emph{et al}. was successfully applied to the hydrogen molecule \cite{burke1986, burke1975}.
The random phase approximation with exchange has been used successfully by Semenov and Cherepkov to the computation of the cross section of the hydrogen \cite{semenov1998} and nitrogen molecules \cite{semenov2000}.
The authors used the zero-order Hartree-Fock numerical basis set representing the discrete and continuum spectrum and solved the RPA equations for the dipole matrix elements with the perturbation given by the correlation potential \cite{amusia1976}.
The advantage of this method is that the results obtained in the length and velocity gauges approximately coincide \cite{semenov1998}.

Recently, the single center method has successfully been applied to $\mathrm{O}_2$, $\mathrm{N}_2$, $\mathrm{NO}$ and $\mathrm{CO}$ molecules \cite{demekhin2011} and also in the description of the circular dichroism \cite{goetz2017}.
This method relies on an old idea of solving the effective one-particle Schr\"{o}dinger equation \cite{bishop1967}.

Yang was the first author who considered scattering angular distribution from an ensemble of randomly oriented species \cite{yang1948}.
From his work, dedicated mostly to nuclear reactions, one concludes that symmetry considerations restrict the form of the differential cross section to~\cite{lin2002}
$$
\pdv{\sigma}{\Omega_{\va{k}}} = \frac{\sigma}{4 \pi} \qty[1 + \beta \, P_2 (\cos \theta^\prime) ],
$$ 
where $\sigma$ is a total cross section and $\beta$ is the so-called asymmetry parameter.
The angle $\theta^\prime$ is the angular distance between asymptotic momentum of photoelectron $\va{k}$ and the polarization vector.
A more elaborate description of the angular distribution was given by Tully~\emph{et~al}.~\cite{tully1968}.

One of the first theoretical works about photoelectron angular distribution in the molecular frame of reference (MFPAD) was published by Dill \cite{dill1976}.
He gave the following formula for the angular distribution in the dipole approximation 
$$
\pdv{\sigma}{\Omega_{k}} = \sum_{l=0}^{2 l_{\mathrm{max}}} \sum_{m=-l}^l A_{lm} Y_{lm} (\theta_k, \phi_k),
$$
where the angles are measured with respect to the molecule-fixed $z$-axis, and $l_{\mathrm{max}}$ is the highest angular momentum component of the photoelectron wavefunction.
The discussion  of the electronic spin projection of the ejected electron was given by Cherepkov and Kuznetsov \cite{cherepkov1987}.
Determination of a minimal set of measurements needed to reproduce any three-dimensional angular distribution is discussed by Semenov and Cherepkov \cite{semenov2009}.
A detailed relation of the MFPAD to the symmetry of the continuum orbital is given by Suzuki \cite{suzuki2008}.

With the development of more advanced experimental techniques, the interest in the molecule-fixed photoelectron angular distribution grows \cite{yagishita2005, yagishita2015}.
There are three main experimental approaches to measure MFPAD \cite{Lucchese2014, reid2003}.
The post-orientation method assumes the axial recoil approximation where molecule is excited to the dissociative state.
Then one determines the molecular orientation from the direction of motion of the ion fragments.
One of the first such experiment was conducted on the $\mathrm{O}_2$ molecule by Golovin, Cherepkov and Kuznetsov \cite{Golovin1992}.
The concurrent-orientation method uses multiple transitions to bound states (exhibiting the alignment dependence) aligning the molecules before the final transition into the continuum.
The third method of pre-orientation, uses the process of aligning the molecules with the help of short infrared laser pulses.
With this method the MFPADs was obtained for $\mathrm{O}_2$, $\mathrm{N}_2$ and $\mathrm{CO}$ \cite{rouzee2012}.

The numerical computation was also used to determine the ionized state wavefunction in the ionization of the $\mathrm{H}_2^+$ molecular ion \cite{picca2005}.
The time dependent density functional theory proved useful in describing the photoionization of $\mathrm{CO}$ molecule \cite{stener2002}.
The random phase approximation with exchange computations of Semenov and Cherepkov \cite{cherepkov2000, semenov2003} for $\mathrm{H}_2$ and $\mathrm{N}_2$ molecules predicted the experimental results from the photoelectron-photoion coincidence method \cite{motoki2002, hikosaka2003}.
Recently, the complex basis set method was used by Matsuzaki and Yaushita in the calculation of MFPADs \cite{matsuzaki2017}.
The authors of this work used a Gaussian basis set with application to the $\mathrm{H}_2$ and $\mathrm{H}_2^+$ molecular systems.

In this paper we introduce a novel approach to the obtaining of the photoionization cross section as well as the photoionization angular distribution.
We base our method on the expansion of the continuum orbital in the square-integrable basis set, with the assumption that the ejected electron  is moving in the Coulomb potential.

The paper is organized as follows.
In Sec.~\ref{sec:model} we present the theoretical model of approximating the Coulomb wavefunctions in a discrete basis set.
Sec.~\ref{sec:results} is devoted to the presentation of the results for H, He and H$_2$ obtained with our model.
Finally we end this paper with the concluding Sec.~\ref{sec:conclusions}.

\section{\label{sec:model} Continuum model}

\subsection{\label{sec:coulomb_wavefunction} Coulomb wavefunction}

In the first order time-dependent perturbation theory the photoionization cross section in the dipole approximation is given by \cite{Bethe}
\begin{equation} \label{eq:cross_section}
\pdv{\sigma}{\Omega_{\va{k}}} = 
\frac{4 \pi ^2 \omega}{c} \,
\abs{\mel{\Psi_{\va{k}}}{\,\va{j} \vdot \va{r}\, }{\Psi_0}}^2.
\end{equation}
Here $\Psi_0$ is the ground state of the system and $\Psi_{\va{k}}$ is the energy-normalized ionized state  of the system with the asymptotic momentum of the photoelectron $\va{k}$.
The ploarization vector is denoted $\va{j}$.
The above formula (\ref{eq:cross_section}) is called the length form of the cross section.
In the case of exact solutions of the Schr\"{o}dinger equation an alternative formula is given by the so-called velocity form
$$
\pdv{\sigma}{\Omega_{\va{k}}} = 
\frac{4 \pi ^2 }{c \, \omega} \,
\abs{\mel{\Psi_{\va{k}}}{\,\va{j} \vdot \va{p}\, }{\Psi_0}}^2.
$$

For the hydrogen atom the state $\Psi_{\va{k}}$ is given by the Coulomb wavefunction $\psi_{\va{k}}^-$ with the incoming spherical wave boundary condition, known analytically \cite{Landau}
\begin{eqnarray} \label{eq:coulomb_wf}
\psi_{\va{k}}^- (\va{r}) =& \sqrt{\frac{k}{(2\pi)^3}} \, e^{i \va{k}\cdot \va{r}} \, \Gamma \qty(1 - i \eta ) \, e^{-\frac{\pi \eta}{2}} \\ \nonumber
&\times \, _1F_1\qty(i \eta, 1, -i \,(kr + \va{k} \cdot \va{r})),
\end{eqnarray}
where $\eta = - Z / k$ is the Sommerfeld parameter, $Z$ the nuclear charge, and $_1F_1$ is the Kummer confluent hypergeometric function \cite{Abramowitz}. 

\subsection{Decomposition} \label{sec:pseudo_partial_waves}

In this section we express (\ref{eq:coulomb_wf}) as
$$
\psi_{\va{k}}^- (\va{r}) = \sqrt{\frac{k}{(2\pi)^3}} \, e^{i \va{k}\cdot \va{r}} 
\sum_{l=0}^\infty \, (kr)^l \,  \mathcal{O}_{kl} (r) \, P_l (\cos \gamma), 
$$
where $P_l$ are the Legendre polynomials and $\gamma$ is an angle between $\va{r}$ and $\va{k}$.
We will use the notation $u \equiv \cos \gamma$.
We call the function $\mathcal{O}_{kl}$ the $l$-th pseudo-partial wave.
In general, we define the overdrive function $\mathcal{O}_{\va{k}} (\va{r})$
$$
\mathcal{O}_{\va{k}} (\va{r}) = 
\sum_{l=0}^\infty \, (kr)^l \,  \mathcal{O}_{kl} (r) \, P_l (\cos \gamma),
$$
as a continuum wavefunction with the plane wave oscillations cut out and seek its partial wave decomposition.
In the general case $\mathcal{O}_{\va{k}} (\va{r})$ satisfies the equation
$$
\qty(-\frac{1}{2} \Delta  + \va{k} \cdot \nabla -  \frac{Z}{r} + v(\va{r}) ) \mathcal{O}_{\va{k}}(\va{r}) = 0,
$$
which is, in general, very difficult to solve analytically.
Here we include the possible deviations from the Coulomb field $v(\va{r})$.
In order to proceed we use the solution (\ref{eq:coulomb_wf}), thus putting $v(\va{r}) = 0$.
From the orthogonality of the Legendre polynomials follows the equation
\begin{eqnarray} \label{eq:o_function_as_integral}
(kr)^l \,  \mathcal{O}_{kl} (r) &&= \frac{2l +1}{2} \, \Gamma \qty(1 - i \eta ) \, e^{-\frac{\pi \eta}{2}} \\ 
&& \times \int_{-1}^1 \dd{u} P_l (u) \, _1F_1\qty(i \eta, 1, -ikr(1+u)). \nonumber
\end{eqnarray}
The last integral is equal to 
$$
\sum_{n=0}^\infty \frac{\qty(i \eta)_n}{(n!)^2} \, (-ikr)^n
\int_{-1}^1 \dd{u} P_l (u) \: (1+u)^n ,
$$
where we used the series representation of the Kummer function~$_1F_1$ and $(i \eta)_n$ is a Pochhammer symbol. 
Let us consider the integral
$$
I_{n l} \equiv \int_{-1}^1 \dd{u} P_l (u) \, (1+u)^n.
$$
From the orthogonality properties of the Legendre polynomials we have the following decomposition
$$
(1 + u)^n = \sum_{s = 0}^n a_s \,P_s(u),
$$
since the left-hand side is a polynomial of degree $n$.
Indeed, $P_s$ with $s \leq n$ form a basis of space of polynomials of degree lower or equal $n$.
Thus, we then have $I_{n l} = 0$ for $ l > n$.
We start the derivation by writing the  Rodrigues' formula for the Legendre polynomials \cite{Abramowitz}
$$
P_l(u) = \frac{1}{2^l \, l!} \, \dv[l]{u} \, (u^2 - 1)^l
$$
and plug it into the integral.
Now, we integrate by parts $l$ times and note that the boundary terms vanish.
We get
$$
I_{n l}=\frac{1}{2^l \, l!} \, \frac{n!}{(n-l)!} \int_{-1}^1 \dd{u} (1-u)^l \, (1+u)^n,
$$
provided that $l \leq n$. 
After the change of variables to $t = (1+u) / 2$ the result reads
$$
\frac{1}{2^l \, l!} \, \frac{n!}{(n-l)!} \, 2^{n+l+1} \int_0^1 \dd{t} t^n \, (1-t)^l.
$$
The last integral coincides with the beta function $\mathrm{B}(n+1, l+1)$~\cite{Abramowitz} and by using the relation 
$$
\mathrm{B}(n+1, l+1) = \frac{\Gamma(n+1) \, \Gamma(l+1)}{\Gamma(n+l+2)}
$$
we can write the integral $I_{n l}$ as
\begin{equation*}
I_{n l} = \frac{2^{n+1} \, (n!)^2}{(l+n+1)! \, (n-l)!}.
\end{equation*}

Going back to Eq. (\ref{eq:o_function_as_integral}), after substitution of the above result we get the equation
\begin{eqnarray} \label{eq:o_function_as_a_sum}
(kr)^l \,  \mathcal{O}_{kl} (r) &&= 
(2l +1) \,  \Gamma \qty(1 - i \eta ) \, e^{-\frac{\pi \eta}{2}} \\ 
&& \times \sum_{n=l}^\infty  (-2ikr)^n
\frac{\qty(i \eta)_n}{(l+n+1)! \, (n-l)!}. \nonumber
\end{eqnarray}
By changing the summation index to $t = n-l$ and using the identities $(t + 2l +1)! = (2l+2)_t \, (2l+1)!$ and $(i\eta)_{t+l} = (i\eta)_l\, (i\eta + l)_t$ we obtain the confluent series
\begin{eqnarray*}
&&\sum_{n=l}^\infty  (-2ikr)^n
\frac{\qty(i \eta)_n}{(l+n+1)! \, (n-l)!} =  \\
&&(-2ikr)^l \, \frac{\qty(i \eta)_l}{(2l+1)!} \, _1F_1 \qty(i \eta + l,2l+2,-2ikr).
\end{eqnarray*}
By combining with Eq. (\ref{eq:o_function_as_a_sum})  we arrive at the radial part of the pseudo-partial wave decomposition
\begin{eqnarray*} 
\mathcal{O}_{kl} (r) =&& \, \Gamma \qty(1 - i \eta ) \, e^{-\frac{\pi \eta}{2}} \qty(i \eta)_l \,\frac{(-2i)^l}{(2l)!} \\
 \times&& \, _1F_1\qty(l + i \eta,  2l+2, -2ikr).
\end{eqnarray*}

\subsection{Asymptotic behavior}\label{sec:ps_waves_asymptocic}

In order to investigate the asymptotic behavior of the pseudo-partial waves we will exploit the theory of the behavior of the Kummer function at infinity \cite{Abramowitz}.
Since $\mathcal{O}_{k l}$ is a function of the product $kr$, we will use the notation $\rho \equiv kr$.
We have the asymptotic expression
\begin{widetext}
\begin{eqnarray*}
\mathcal{O}_{kl} &&\rightarrow
 \frac{\Gamma(1 - i \eta)}{\Gamma(i \eta)} \, (2l+1) \, (- \rho)^{-l} \, e^{-i \eta \log 2\rho} \, e^{2 i \arg \Gamma(l + i \eta)} \, \times 
 \Bigg\{
 \frac{1}{(l+1-i\eta) \, (l - i \eta)} 
 - \frac{l+i\eta}{l - i\eta} \, (2 i \rho)^{-1} + \\ 
 &&\qty[ \frac{(l+1+i\eta) \, (l + i\eta)}{2}  + 
 e^{-2i \rho} \, e^{2 i \eta \log 2\rho} \, e^{-2 i \arg \Gamma(l + i \eta)} \, e^{i \pi l}
 \, ]
  (2 i \rho)^{-2} + \order{\rho^{-3}}
   \Bigg\}.
  \end{eqnarray*}
\end{widetext}
The oscillatory term $e^{-2i \rho}$ appears only in the order $\order{\rho^{-2}}$ and higher in the curly brackets.
Thus, $\mathcal{O}_{kl}$ are much smoother than the ordinary partial waves $R_{k l}$ which behave asymptotically as \cite{Landau}
\begin{eqnarray*}
R_{k l}(r) &&\rightarrow  \sqrt{\frac{2}{\pi}} \, \frac{1}{r} \\
&&\times \sin(kr - \eta \ln 2kr -\frac{\pi l}{2} + \arg \Gamma(l+1+i\eta)),
\end{eqnarray*}
and the oscillatory terms appear already in the leading order.

\subsection{\label{sec:basis_set} Basis set decomposition}

Let us note that in the photoionization process we do not need to consider the continuum orbital at an arbitrary distance from the ion.
All necessary matrix elements involve at least one orbital with a bound character that decays exponentially.
Thus it is justified to put large distance cutoff on the continuum orbital, thereby representing it in an $L^2$ basis set as follows.
We expand the function $\mathcal{O}_{kl} (r)$ in a basis of the GTOs obtaining the approximate expression
$$
\mathcal{O}_{kl} (r) \approx \sum_{i=1}^N c_{l i} \, e^{-\xi_{l i} r^2}.
$$
The GTO basis is parameterized by the exponents $( \xi_{l i} )_{i = 1, \dots N}$.
From the computational point of view it is convenient to introduce the regular solid harmonics
$$
\mathcal{R}_{lm} (\va{r}) = \sqrt{\frac{4 \pi}{2l+1}} r^l Y_{lm}(\hat{r}),
$$
which are harmonic and homogeneous polynomials of $(x, y, z)$ of degree $l$.
The Coulomb wavefunction can then be approximated as 
\begin{equation} \label{eq:coulomb_wf_approximated}
\psi_{\va{k}}^- (\va{r}) \approx
\sum_{l=0}^\infty
\sum_{m=-l}^l 
\sum_{i=1}^N \, 
C_{lm i} (\va{k}) \: e^{i \va{k}\cdot \va{r}} \, e^{-\xi_{l i} r^2} \mathcal{R}_{lm} (\va{r}),
\end{equation}
where $C_{lm i}$ are the new expansion coefficient and read
$$
C_{lm i} (\va{k}) = (2\pi)^{-\frac{3}{2}} \sqrt{k} \, c_{l i} \, \mathcal{R}_{lm}^\star (\va{k}).
$$
In actual computations the sum over $l$ in Eq. (\ref{eq:coulomb_wf_approximated}) must be truncated for practical reasons and this issue is discussed further in the text.

Now we see that we implicitly decomposed the continuum Coulomb wavefunction into the Gaussian type orbitals times the plane waves (GTOPW) of the form
\begin{eqnarray} \label{eq:gtopw}
&&\phi_{n_x n_y n_z} (\va{k}, \xi, \va{A}; \va{r}) = e^{i \va{k}\cdot (\va{r}-\va{A})} e^{-\xi (\va{r}-\va{A})^2} \\
&& \qquad \times 
(x-A_x)^{n_x} \, (y-A_y)^{n_y} \, (z-A_z)^{n_z}, \nonumber
\end{eqnarray}
where $\va{A}$ is the centering point. Some authors refer to them as the London orbitals \cite{Tellgren2008}.
Previously they were used in the calculations of molecular properties in strong magnetic fields \cite{Tellgren2012}.
Same as for the GTOs, for the GTOPWs we have the product theorem stating that the product of two GTOPWs is a linear combination of GTOPWs. 
For the polynomial part in Eq. (\ref{eq:gtopw}) this result is obvious, so let us focus on the exponential part.
We have 
\begin{eqnarray*}
&&e^{-\xi_A (\va{r} - \va{A})^2} \, e^{i \va{k}_A \vdot (\va{r} - \va{A})} \,
e^{-\xi_B (\va{r} - \va{B})^2} \, e^{i \va{k}_B \vdot (\va{r} - \va{B})}
= \\
&&e^{-i (\va{k}_A \vdot \va{A} + \va{k}_B \vdot \va{B})} \,
e^{i \va{k}_P \vdot \va{P}} \,
e^{-\xi_P (\va{r} - \va{P})^2} \, e^{i \va{k}_P \vdot (\va{r} - \va{P})},
\end{eqnarray*}
where $\xi_P = \xi_A + \xi_B$, $\va{k}_P = \va{k}_A + \va{k}_B$ and the new centering point is $\va{P} = (\xi_A \va{A} + \xi_B \va{B}) / \xi_P$.
Computation of one- and two-electron integrals with the GTOPWs has recently been discussed in the literature in details \cite{irons17}. 
In our implementation all necessary integrals are computed by using a generalization of the McMurchie-Davidson scheme~\cite{Tachikawa2001}.
Let us also mention that in computations we used the \textsc{Eigen} linear algebra library \cite{eigen} for manipulations on matrices and vectors.

\subsection{\label{sec:optimization_pwgto} Optimization of GTOPW}

Let us describe the process of approximating the Coulomb wavefunction by the linear combination of GTOPWs.
We chose to approximate one pseudo-partial wave by a contraction of $n=10$ GTOPW orbitals.
The plane wave part of the contraction is specified by the vector $\va{k}$ and for the exponents of the Gaussian functions we use the well-tempering method \cite{huzinaga1990}.
In this method we assume that the exponents $(\xi_k)_{k=1,\dots, n}$ are given by the formula  
$$
\xi_k = \alpha \, \beta^{k-1} \qty(1 + \gamma \qty(\frac{k-1}{n})^\delta).
$$
Thus, we reduced the optimization problem to finding the optimal values of the parameters $(\alpha, \beta, \gamma, \delta)$.

Optimization of the exponents is done by producing tabulated values of the pseudo-partial wave $\mathcal{O}_{kl}(r_i)$ on a radial grid $r_i$, extending over the domain from 0 to 30~a.u. with spacing $\Delta r = 0.1$~a.u.,  and minimizing the expression
$$
R^2 = \sum_i \qty(\mathcal{O}_{kl}(r_i) - \sum_k c_{l k} \, e^{-\xi_{l k} r_i^2})^2.
$$
For a given set of the parameters $(\alpha, \beta, \gamma, \delta)$ the value of $R^2$ and of the linear coefficients $c_{l k}$ are found by the least-squares method. Therefore, the non-linear optimization is necessary only for the parameters $(\alpha, \beta, \gamma, \delta)$.

We chose to optimize this problem in two steps.
The first optimization is done by means of the evolutionary algorithm \cite{Eiben}.
This method mimics the biological evolution where a fitness value is assigned to a genotype which gives the odds for participating in the creation of the next generation.
In our case the genotype is a 4-component vector of parameters $(\alpha, \beta, \gamma, \delta)$.
We use the evolutionary method because we found that our minimization problem consist of a large number of local minima.
Our method, with a proper adjustment of crossover and mutation rates, is superior in finding an approximate global minimum compared to the standard algorithms like BFGS \cite{fletcher}.
The evolutionary algorithm gives an approximate global minimum which can be improved.
In order to achieve it we use the Levenberg-Marquardt damped least-squares method \cite{fletcher} with the starting point given by the evolutionary algorithm.
The Levenberg-Marquardt algorithm finds the minimum closest to the starting point in a deterministic way, thus improving the results.

\begin{figure*}
\centering
\begin{tabular}{c c}
\includegraphics[width=0.49\textwidth]{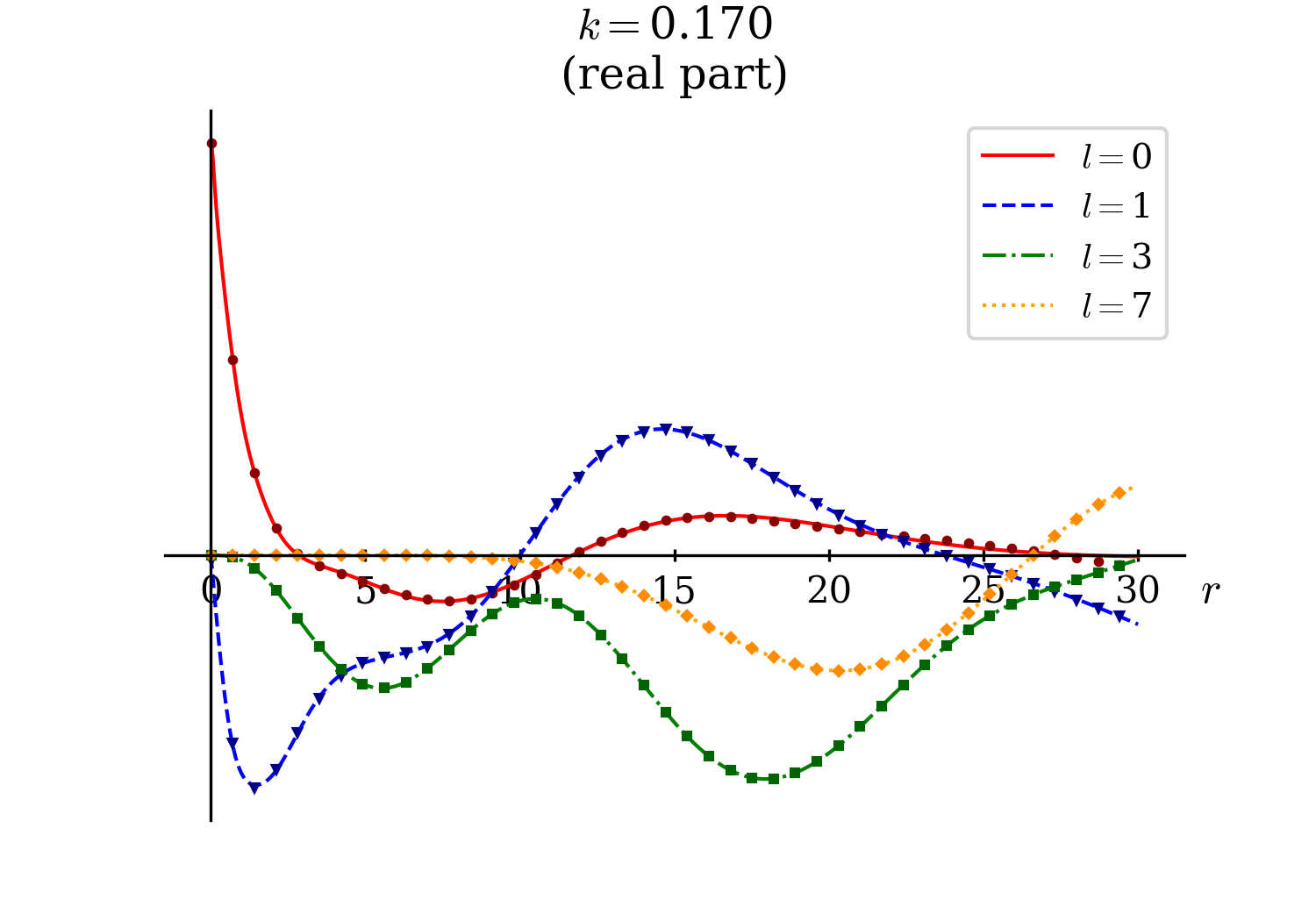} & 
\includegraphics[width=0.49\textwidth]{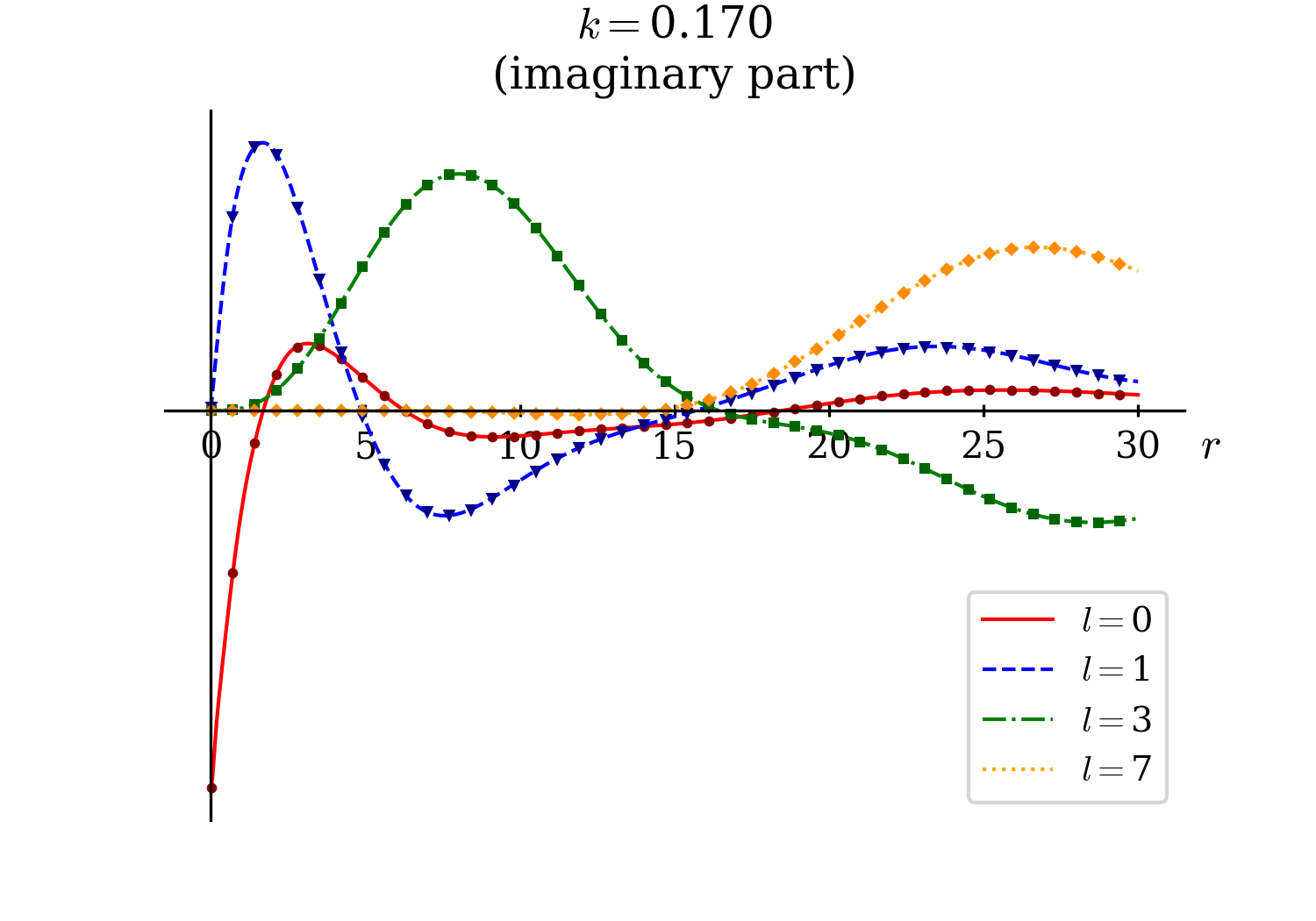} \\
\includegraphics[width=0.49\textwidth]{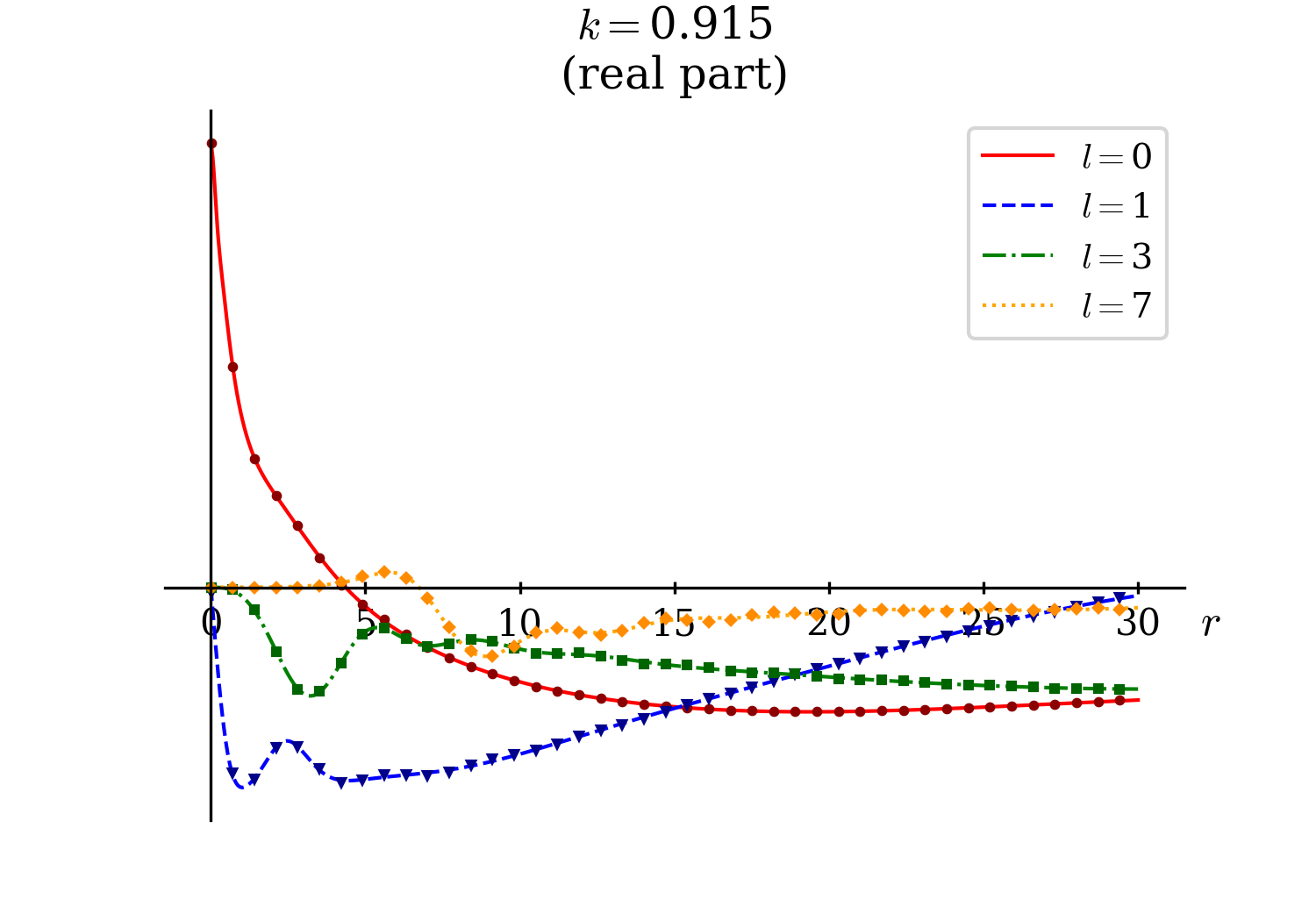} & 
\includegraphics[width=0.49\textwidth]{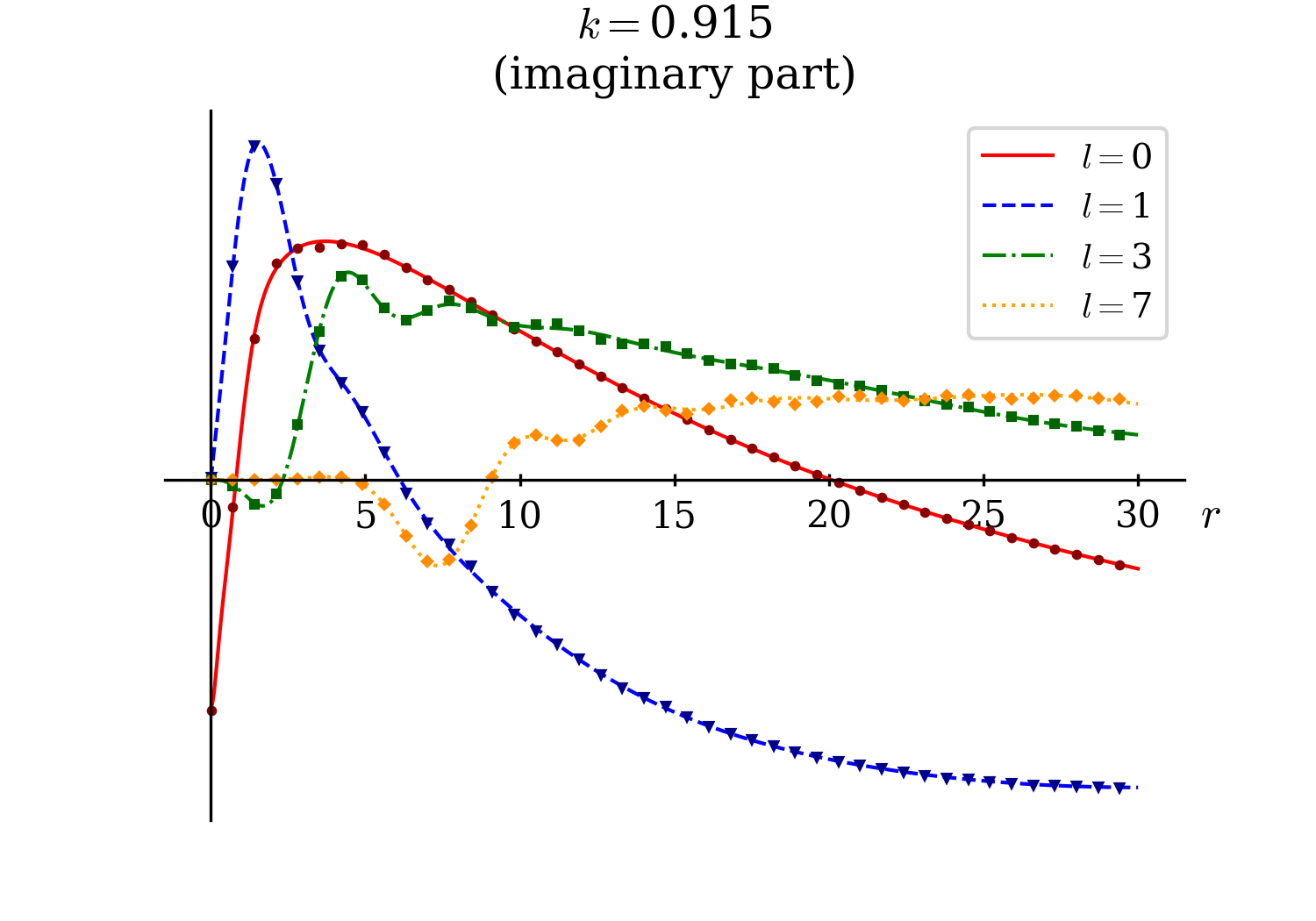} \\
\includegraphics[width=0.49\textwidth]{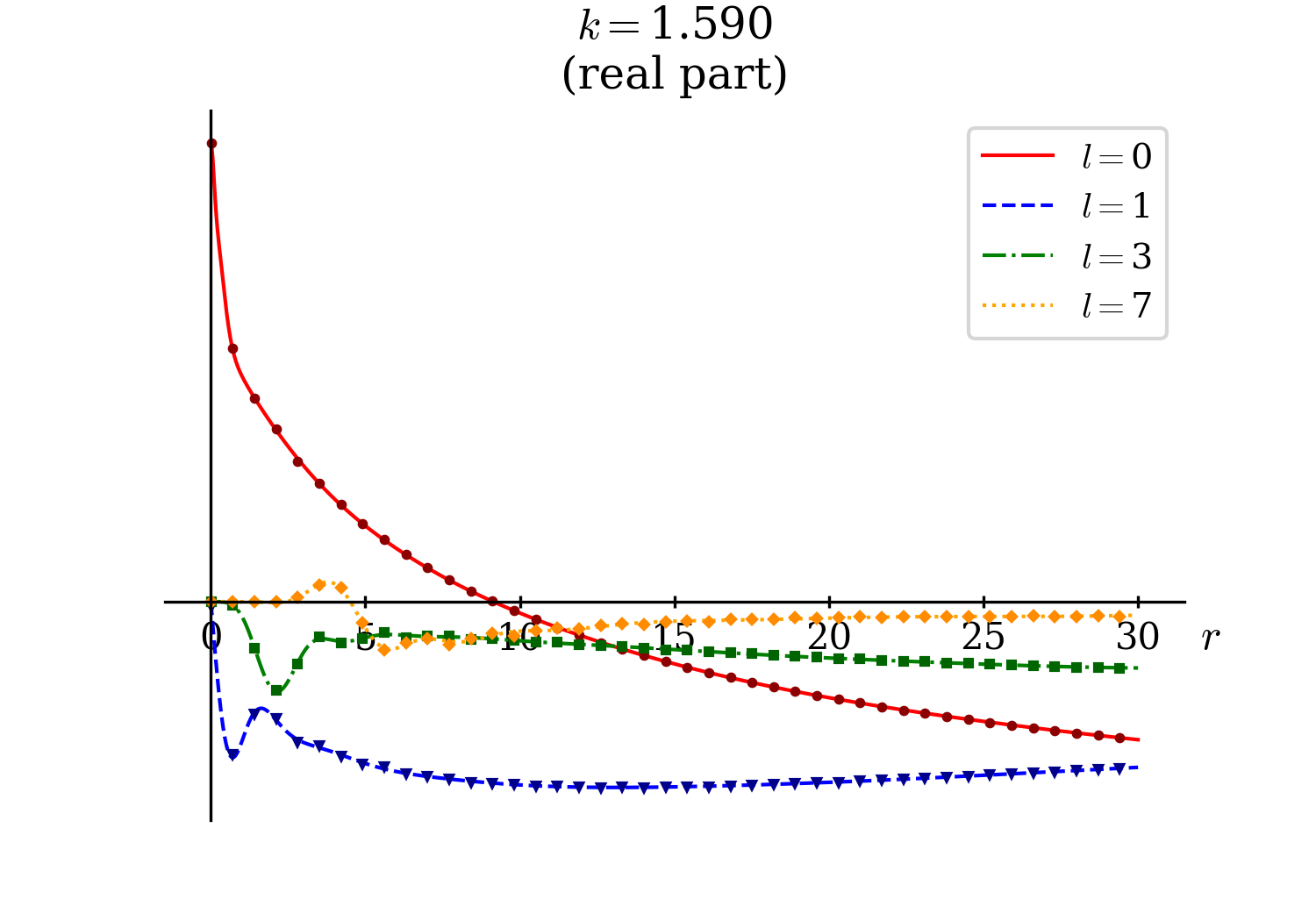} & 
\includegraphics[width=0.49\textwidth]{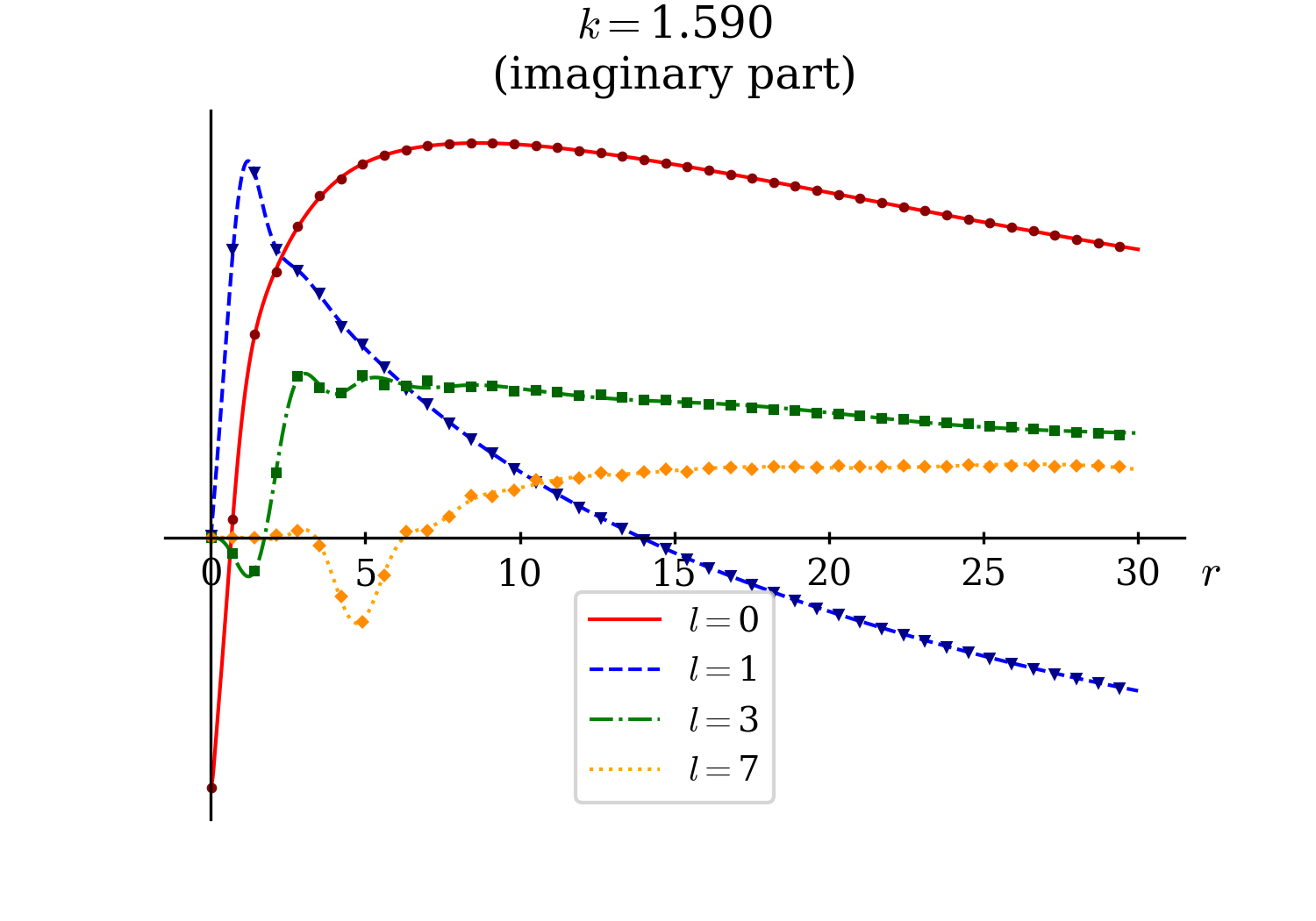}
 \end{tabular}
\caption{Examples of obtained fits of pseudo-partial waves $(kr)^l \, \mathcal{O}_{kl}$, represented by lines. The tabulated reference values of pseudo-partial waves given in solid marks. Radial distance $r$ given in atomic units.}
\label{fig:fits}
\end{figure*}

The examples of obtained fits are presented on Figure~\ref{fig:fits}.

\subsection{\label{sec:construction} Construction of states} 

Let us focus on the two-electrons systems.
We assume that the final and initial states are both spin singlets.
This assumption is unnecessary but simplifies the discussion.
The interaction Hamiltonian does not act on the spin part of the wavefunction, thus in the absence of the spin-orbit interaction the time evolution does not mix spin states.
We construct the final state electronic wavefunction as
\begin{equation} \label{eq:final_state}
\Psi_{f} ( \va{r}_1, \va{r}_2 ) =
\frac{1}{\sqrt{2}} \, \qty( \psi_{\va{k}} (\va{r}_1) \, \psi_0 (\va{r}_2) +
\psi_{\va{k}} (\va{r}_2) \, \psi_0 (\va{r}_1)),
\end{equation}
where $\psi_0$ denotes the ion orbital and $\psi_{\va{k}}$ is the continuum orbital.

We assume that $\psi_0$ decays exponentially, whereas $\psi_{\va{k}}$ has the asymptotic form of the Coulomb wavefunction. 
This follows from the fact that the deviations of the potential in which ejected electron moves decays faster than the Coulomb potential itself.
We want $\Psi_f$ to satisfy the Schr\"{o}dinger equation 
\begin{equation} \label{eq:eigenvalue_problem}
H \Psi_f = E \Psi_f .
\end{equation}
The Hamiltonian is of the standard form 
$$
H  = h_1 + h_2 + V_{12},
$$
where $h_1$ and $h_2$ are one-electron operators acting on functions of $\va{r}_1$ and $\va{r}_2$, respectively. 
The electron repulsion potential $V_{12}$ is a two-electron operator acting on both variables.

Let us insert $\Psi_f$ into Eq. (\ref{eq:eigenvalue_problem}), multiply by $\psi_0 (\va{r}_2)$ from the left and integrate over $\va{r}_2$. The result reads
\begin{widetext}
\begin{eqnarray*}
&&\ip*{\psi_0}{\psi_{\va{k}}} \, h_1 \, \psi_0 (\va{r}_1) \,+ \,
\mel*{\psi_0}{h}{\psi_{\va{k}}} \, \psi_0 (\va{r}_1) \,+ \,
\qty( \int \dd{\va{r}_2} \psi_0^\star (\va{r}_2) \, V_{12} \, \psi_{\va{k}} (\va{r}_2) ) 
\, \psi_0 (\va{r}_1) \, + \\
&&h_1 \, \psi_{\va{k}} (\va{r}_1) \,+ \,
\mel*{\psi_0}{h}{\psi_0}\, \psi_{\va{k}} (\va{r}_1) \,+ \,
\qty( \int \dd{\va{r}_2} \psi_0^\star (\va{r}_2) \, V_{12} \, \psi_0 (\va{r}_2) ) \, \psi_{\va{k}} (\va{r}_1) \, = \\
&&E \, \qty( \ip*{\psi_0}{\psi_{\va{k}}} \, \psi_0 (\va{r}_1) \,+ \,
\psi_{\va{k}} (\va{r}_1) ),
\end{eqnarray*}
\end{widetext}
where $\ip{\psi_0}{\psi_0} = 1$ has been used.

The above equation can be a subject to many approximations.
The same equation was used by Stewart \emph{et al}. to compute the total photoionization cross section of the He atom and $\mathrm{Li}^+$ ion \cite{stewart1960, stewart1963}.
In our computations, however, we used the simplest approximation, namely $\psi_{\va{k}}$ is the Coulomb wavefunction and $\psi_0$ is the lowest energy electronic wavefunction of the ion.

Let us also discuss the construction of the initial state $\Psi_i$.
In the Hartree-Fock approximation one can take $\Psi_i$ as a single determinant
$$
\Psi_i (\va{r}_1, \va{r}_2) = \varphi_0 (\va{r}_1) \, \varphi_0 (\va{r}_2).
$$
More accurate results can be obtained by employing the CI expansion in terms of the reference orbitals $\varphi_i$ obtained from the Hartree-Fock method
$$
\Psi_i (\va{r}_1, \va{r}_2) = \sum_{i,j} c^{ij} \, \varphi_i (\va{r}_1) \, \varphi_j (\va{r}_2).
$$
Due to the symmetrization of the spatial part of the total wavefunction the coefficients $c^{ij}$ remain unchanged under the transposition of the indices.
They also must be set in such a way that the whole state is normalized to the unity.

One of the major advantages of the approach based on the GTOPW representation of the continuum orbitals is the fact it can be extended to many-electron atoms and molecules without difficulties encountered in some other methods. This claim is based on two observations. First, the matrix elements involving the GTOPWs can be evaluated for an arbitrary geometry of a molecular system - a task that is very difficult and/or computationally demanding, e.g., in the grid-based methods. Second, the construction of states for many-electron systems can be accomplished by using the coupled-cluster (CC) \cite{bartlett07} and equation-of-motion (EOM) CC \cite{monkhorst77,sekino84,geertsen89,stanton93} representations of the electronic wavefunctions. In the case of the initial state the usual CC exponential parametrization can be adopted, $\Psi_i=e^T\Phi_0$, where $T$ is the cluster operator and $\Phi_0$ is the reference determinant. The simplest approximation to the final state is obtained by employing EOM-CC form, namely $\Psi_f=\Omega\,e^T\Phi_0$, where $\Omega$ is an excitation operator. Note that in the photoionization problem the operator $\Omega$ must possess a specific form where one of the excitations always leads to the prepared continuum orbital. The remaining elements of this excitation operator are calculated by diagonalizing the EOM-CC Jacobian matrix (which is complex non-Hermitian in this case). Finally, the transition moments required to calculate the photoionization cross-section, cf. Eq. (\ref{eq:cross_section}), are obtained from rather straightforward generalizations of the linear-response CC theory \cite{koch90,koch94,pedersen97} or the time-independent XCC theory \cite{moszynski05,tucholska14,tucholska17}.

\section{\label{sec:results} Results}

\subsection{\label{sec:hydrogen} Hydrogen atom}

In order to test the method presented in Sec.~\ref{sec:model} we performed computations of the total ionization cross section for the hydrogen atom.
The exact result was obtained by Stobbe in 1930  \cite{stobbe1930} and later by Menzel and Perkins \cite{menzel1935}.
There is also a limited set of the experimental data \cite{beynon1965}. 
However, we decided to compare our results with reference theoretical results obtained numerically from what follows.

The initial state is given by the lowest energy state of hydrogen atom $\psi_{100}$ with the zero angular momentum and energy $E_0 = -0.5$~a.u.
The final state is the Coulomb continuum wavefunction $\psi_{\va{k}}^-$ with the energy $E = k^2 / 2$.
By standard manipulations we easily simplify the Eq.~(\ref{eq:cross_section}) for the hydrogen atom obtaining 
\begin{equation} \label{eq:h_atom_exact}
\sigma = \frac{4 \pi^2 \omega}{c \, k} 
\qty[ \int_0^\infty \dd{r} r^3 \, R_{k1}(r) \, \psi_{100} (r) ]^2,
\end{equation}
where $R_{k1}$ is a Coulomb partial wave with the normalization
$$
\int_0^\infty \dd{r} r^2 R_{k^\prime l }(r) \, R_{k l} (r) = \delta(k^\prime - k).
$$
We also note that the energy conservation is implied $\omega + E_0 = E$.
The formula (\ref{eq:h_atom_exact}) is well suited to obtain the reference cross section values.

To test out method we represented the initial ground state wavefunction $\psi_{100}$ in a basis of 9 GTOs.
The solution was obtained variationally with the help of the \textsc{GAMESS} package \cite{gamess}.
The value of the variational energy of the ground state was $E = -0.4999968$~a.u.
The Coulomb wavefunction $\psi_{\va{k}}^-$ was represented as a sum of pseudo-partial waves truncated at $l=8$.
We also performed the computation with the truncation at $l=6$ in order to see the convergence of the pseudo-partial waves series.
Details of the pseudo-partial wave decomposition are presented in Sec. \ref{sec:pseudo_partial_waves}.
Each pseudo-partial wave $\mathcal{O}_{kl}$ was represented as a linear combination of 10 Gaussian functions.
We used the following approximation for the Coulomb wavefunction
\begin{eqnarray}
\label{eq:approx_coulomb_wf}
\psi_{\va{k}}^- (\va{r}) \approx&& 
\sum_{l=0}^8 \, \sum_{j=1}^{10} \, \sum_{n_x + n_y +n_z = l} C_{n_x n_y n_z, j} (\va{k}) \nonumber \\
&&\times \phi_{n_x n_y n_z} (\va{k}, \xi_{l j}, \va{0}; \va{r}),
\end{eqnarray}
where $\phi_{n_x n_y n_z}$ is a GTOPW given by (\ref{eq:gtopw}) and the sum over $l$ is truncated.
The position of the nucleus is chosen as the origin $\va{0}$.
Details of the fitting procedure are described in Sec. \ref{sec:optimization_pwgto}.

\begin{figure*}
\begin{subfigure}{0.48\textwidth}
   \centering
    \includegraphics[width=\textwidth]{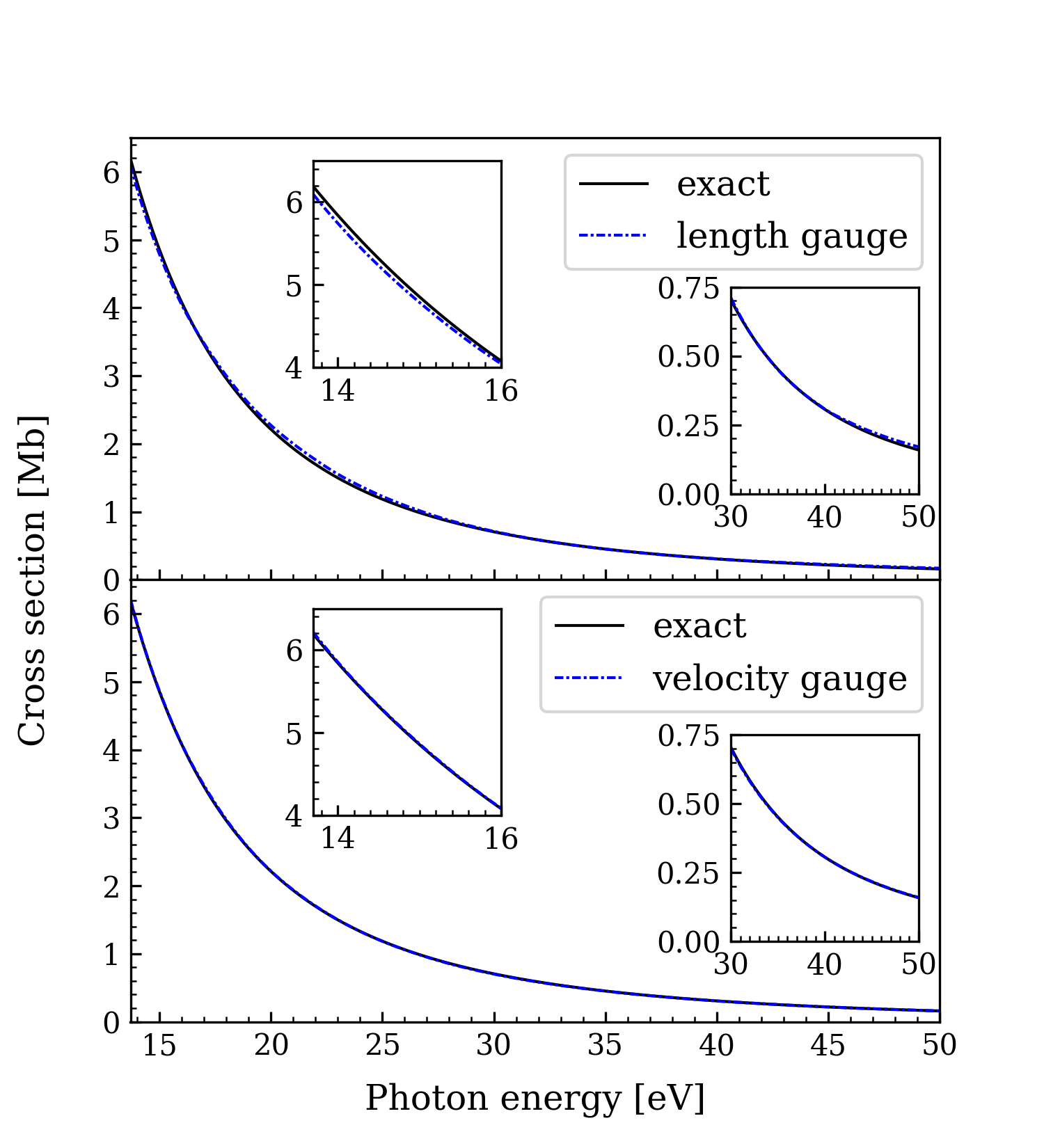}
    \caption{The results with the truncation at $l=8$.}
    \label{fig:atom_h_8}
\end{subfigure}
\begin{subfigure}{0.48\textwidth}
   \centering
    \includegraphics[width=\textwidth]{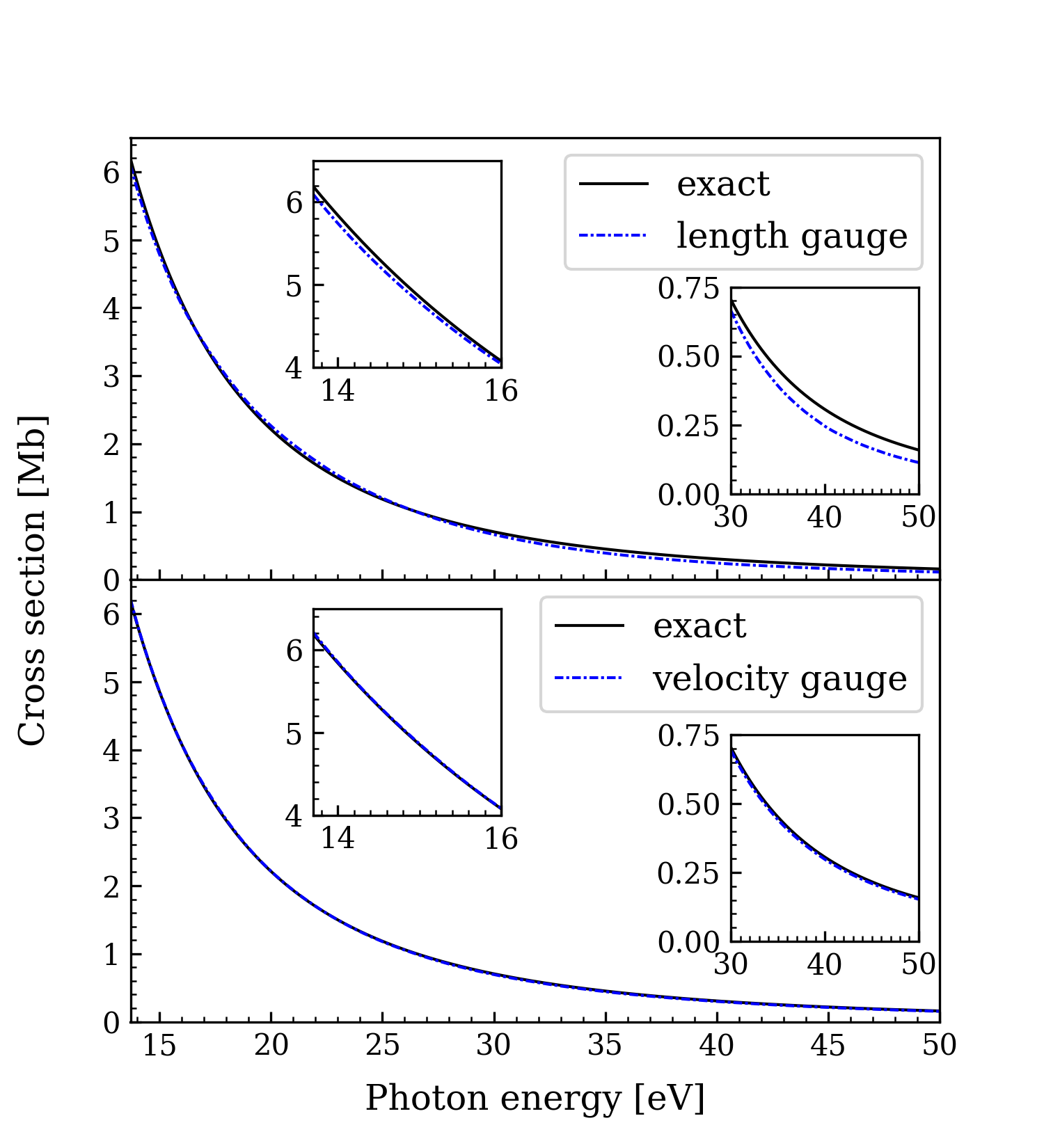}
    \caption{The results with the truncation at $l=6$.}
    \label{fig:atom_h_6}
\end{subfigure}
\caption{The photoionization cross section of the hydrogen atom. The exact results were obtained by numerical integration. The lines were obtained as interpolation of the results from Tables~\ref{tab:atom_h_6} and~\ref{tab:atom_h_8}. }
\end{figure*}

The computations were preformed with the length and velocity representations of the expression for the cross section.
The exact solutions were obtained by the numerical integration of (\ref{eq:h_atom_exact}) using the \textsc{Mathematica} package \cite{mathematica}.
Numerical results are presented in Tables \ref{tab:atom_h_8} and \ref{tab:atom_h_6}.
The naming convention is as follows: $\sigma_{\mathrm{ex}}$ is an exact cross section, $\sigma_{\mathrm{l}}$ and $\sigma_{\mathrm{v}}$ are cross sections computed using the length and velocity gauges, respectively.
The error of the computed result with respect to the exact one is also reported.
The interpolated data are visualized in Figures \ref{fig:atom_h_8} and \ref{fig:atom_h_6}.

\begin{table}
\centering
\caption{Photoionization cross section for the hydrogen atom as a function of energy with the truncation at $l=8$. The relative error is denoted $\bar{\sigma}$. } \label{tab:atom_h_8}
\begin{ruledtabular}
\squeezetable
\begin{tabular}{d d d d d d d}
\hline
\multicolumn{1}{c}{$\omega$~[eV]} & 
\multicolumn{1}{c}{$k$~[au]} & 
\multicolumn{1}{c}{$\sigma_{\mathrm{ex}}$~[Mb]} & 
\multicolumn{1}{c}{$\sigma_{\mathrm{l}}$~[Mb]} &
\multicolumn{1}{c}{$\sigma_{\mathrm{v}}$~[Mb]} &
\multicolumn{1}{c}{$\bar{\sigma}_{\mathrm{l}}$~[\%]} & 
\multicolumn{1}{c}{$\bar{\sigma}_{\mathrm{v}}$~[\%]} \\
\hline
13.7 & 0.083 & 6.189 & 6.094 & 6.213 & 1.54 & 0.38 \\
14.0 & 0.170 & 5.841 & 5.744 & 5.852 & 1.66 & 0.19 \\
15.0 & 0.320 & 4.854 & 4.787 & 4.862 & 1.39 & 0.17 \\
16.0 & 0.419 & 4.077 & 4.043 & 4.082 & 0.84 & 0.10 \\
17.0 & 0.499 & 3.458 & 3.476 & 3.475 & 0.54 & 0.49 \\
18.0 & 0.568 & 2.957 & 2.998 & 2.971 & 1.37 & 0.46 \\
19.0 & 0.630 & 2.549 & 2.596 & 2.556 & 1.85 & 0.27 \\
20.0 & 0.686 & 2.212 & 2.272 & 2.216 & 2.71 & 0.18 \\
21.0 & 0.737 & 1.932 & 2.000 & 1.938 & 3.55 & 0.30 \\
22.0 & 0.785 & 1.697 & 1.764 & 1.702 & 3.97 & 0.29 \\
23.0 & 0.831 & 1.498 & 1.554 & 1.500 & 3.70 & 0.13 \\
24.0 & 0.874 & 1.329 & 1.379 & 1.329 & 3.75 & 0.04 \\
25.0 & 0.915 & 1.185 & 1.226 & 1.183 & 3.54 & 0.13 \\
26.0 & 0.954 & 1.060 & 1.094 & 1.058 & 3.23 & 0.14 \\
27.0 & 0.992 & 0.952 & 0.978 & 0.949 & 2.68 & 0.30 \\
28.0 & 1.029 & 0.859 & 0.874 & 0.854 & 1.84 & 0.55 \\
29.0 & 1.064 & 0.777 & 0.788 & 0.772 & 1.45 & 0.55 \\
30.0 & 1.098 & 0.705 & 0.711 & 0.701 & 0.95 & 0.58 \\
31.0 & 1.131 & 0.641 & 0.645 & 0.638 & 0.57 & 0.57 \\
32.0 & 1.163 & 0.585 & 0.586 & 0.582 & 0.13 & 0.55 \\
33.0 & 1.194 & 0.536 & 0.535 & 0.533 & 0.04 & 0.48 \\
34.0 & 1.224 & 0.491 & 0.491 & 0.490 & 0.03 & 0.34 \\
35.0 & 1.254 & 0.452 & 0.451 & 0.450 & 0.26 & 0.37 \\
36.0 & 1.283 & 0.416 & 0.415 & 0.415 & 0.23 & 0.31 \\
37.0 & 1.311 & 0.384 & 0.385 & 0.384 & 0.05 & 0.14 \\
38.0 & 1.339 & 0.355 & 0.356 & 0.355 & 0.17 & 0.12 \\
39.0 & 1.366 & 0.330 & 0.331 & 0.330 & 0.58 & 0.02 \\
40.0 & 1.393 & 0.306 & 0.307 & 0.306 & 0.38 & 0.05 \\
41.0 & 1.419 & 0.285 & 0.287 & 0.285 & 0.78 & 0.09 \\
42.0 & 1.445 & 0.265 & 0.270 & 0.266 & 1.75 & 0.11 \\
43.0 & 1.470 & 0.247 & 0.253 & 0.248 & 2.41 & 0.23 \\
44.0 & 1.495 & 0.231 & 0.238 & 0.232 & 3.00 & 0.29 \\
45.0 & 1.519 & 0.216 & 0.225 & 0.217 & 3.82 & 0.44 \\
46.0 & 1.543 & 0.203 & 0.212 & 0.204 & 4.49 & 0.49 \\
47.0 & 1.567 & 0.190 & 0.200 & 0.191 & 5.02 & 0.45 \\
48.0 & 1.590 & 0.179 & 0.189 & 0.180 & 5.82 & 0.54 \\
49.0 & 1.613 & 0.168 & 0.179 & 0.169 & 6.46 & 0.55 \\
50.0 & 1.636 & 0.158 & 0.170 & 0.159 & 7.28 & 0.52 \\
\end{tabular}
\end{ruledtabular}
\end{table}
\begin{table}
\centering
\caption{Photoionization cross section for the hydrogen atom as a function of energy with the truncation at $l=6$. The relative error is denoted $\bar{\sigma}$. } \label{tab:atom_h_6}
\begin{ruledtabular}
\squeezetable
\begin{tabular}{d d d d d d d}
\hline
\multicolumn{1}{c}{$\omega$~[eV]} & 
\multicolumn{1}{c}{$k$~[au]} & 
\multicolumn{1}{c}{$\sigma_{\mathrm{ex}}$~[Mb]} & 
\multicolumn{1}{c}{$\sigma_{\mathrm{l}}$~[Mb]} &
\multicolumn{1}{c}{$\sigma_{\mathrm{v}}$~[Mb]} &
\multicolumn{1}{c}{$\bar{\sigma}_{\mathrm{l}}$~[\%]} & 
\multicolumn{1}{c}{$\bar{\sigma}_{\mathrm{v}}$~[\%]} \\
\hline
13.7 & 0.083 & 6.189 & 6.094 & 6.213 & 1.54  & 0.38  \\
14.0 & 0.170 & 5.841 & 5.744 & 5.852 & 1.66  & 0.19  \\
15.0 & 0.320 & 4.854 & 4.787 & 4.862 & 1.39  & 0.17  \\
16.0 & 0.419 & 4.077 & 4.042 & 4.081 & 0.86  & 0.10  \\
17.0 & 0.499 & 3.458 & 3.474 & 3.474 & 0.48  & 0.48  \\
18.0 & 0.568 & 2.957 & 2.994 & 2.970 & 1.25  & 0.43  \\
19.0 & 0.630 & 2.549 & 2.590 & 2.555 & 1.63  & 0.22  \\
20.0 & 0.686 & 2.212 & 2.264 & 2.214 & 2.36  & 0.11  \\
21.0 & 0.737 & 1.932 & 1.990 & 1.936 & 3.02  & 0.20  \\
22.0 & 0.785 & 1.697 & 1.751 & 1.699 & 3.18  & 0.14  \\
23.0 & 0.831 & 1.498 & 1.536 & 1.497 & 2.56  & 0.08  \\
24.0 & 0.874 & 1.329 & 1.358 & 1.325 & 2.15  & 0.34  \\
25.0 & 0.915 & 1.185 & 1.201 & 1.178 & 1.37  & 0.52  \\
26.0 & 0.954 & 1.060 & 1.064 & 1.053 & 0.38  & 0.63  \\
27.0 & 0.992 & 0.952 & 0.943 & 0.944 & 0.95  & 0.91  \\
28.0 & 1.029 & 0.859 & 0.836 & 0.848 & 2.69  & 1.28  \\
29.0 & 1.064 & 0.777 & 0.745 & 0.766 & 4.07  & 1.42  \\
30.0 & 1.098 & 0.705 & 0.665 & 0.694 & 5.65  & 1.60  \\
31.0 & 1.131 & 0.641 & 0.595 & 0.630 & 7.19  & 1.74  \\
32.0 & 1.163 & 0.585 & 0.534 & 0.574 & 8.85  & 1.87  \\
33.0 & 1.194 & 0.536 & 0.480 & 0.525 & 10.30 &  1.96 \\
34.0 & 1.224 & 0.491 & 0.434 & 0.482 & 11.62 &  1.97 \\
35.0 & 1.254 & 0.452 & 0.392 & 0.442 & 13.23 &  2.17 \\
36.0 & 1.283 & 0.416 & 0.355 & 0.407 & 14.61 &  2.27 \\
37.0 & 1.311 & 0.384 & 0.324 & 0.376 & 15.78 &  2.27 \\
38.0 & 1.339 & 0.355 & 0.295 & 0.347 & 17.12 &  2.40 \\
39.0 & 1.366 & 0.330 & 0.269 & 0.321 & 18.22 &  2.43 \\
40.0 & 1.393 & 0.306 & 0.245 & 0.298 & 19.88 &  2.65 \\
41.0 & 1.419 & 0.285 & 0.225 & 0.277 & 20.99 &  2.68 \\
42.0 & 1.445 & 0.265 & 0.208 & 0.258 & 21.63 &  2.81 \\
43.0 & 1.470 & 0.247 & 0.192 & 0.240 & 22.52 &  2.85 \\
44.0 & 1.495 & 0.231 & 0.177 & 0.224 & 23.49 &  2.94 \\
45.0 & 1.519 & 0.216 & 0.164 & 0.210 & 24.27 &  2.95 \\
46.0 & 1.543 & 0.203 & 0.152 & 0.197 & 25.17 &  3.05 \\
47.0 & 1.567 & 0.190 & 0.140 & 0.184 & 26.19 &  3.24 \\
48.0 & 1.590 & 0.179 & 0.131 & 0.173 & 26.99 &  3.29 \\
49.0 & 1.613 & 0.168 & 0.121 & 0.162 & 27.92 &  3.43 \\
50.0 & 1.636 & 0.158 & 0.113 & 0.153 & 28.69 &  3.60 \\
\end{tabular}
\end{ruledtabular}
\end{table}

From the presented results for the hydrogen atom follows that our model exhibits excellent agreement with exact results.
The truncation at $l=8$ suffices to reliably represent the Coulomb wavefunction in considered range of energies.
We note that in the case of truncation at $l=6$ the values of cross section start visibly differ at Fig.~\ref{fig:atom_h_6} from the exact one withing the length gauge for energies higher than 30~eV.

\subsection{Helium atom} \label{sec:helium_atom}

The theoretical values of the helium photoionization cross section are important when studying the behavior of the interstellar matter subjected to radiation, thus they are of great importance in astrophysical computations \cite{yan1998}. 
The early theoretical investigation began with the use of Coulomb wavefunction for the continuum orbital and bound orbitals derived from Ritz variational principle  \cite{huang1948}.
There was an attempt to compute the continuum wavefunction numerically \cite{stewart1960, stewart1963} from the equations derived in Sec. \ref{sec:construction} and a good agreement with the experimental values was obtained.
The extension of this method, using the Hartree-Fock determinant as the final wavefunction was also investigated for the ions from the helium isoelectronic series~\cite{bell1971}.
The use of B-spline basis set as a representation of the continuum wave was reported \cite{venuti1996}, also in the resonance region between 59 eV and the threshold energy for the $2s$ channel at 68.4~eV.
There is also a recent work employing the complex optimized GTOs basis method \cite{morita2008}.

Computationally the topic of photoionization of the helium atom is heavily exploited.
We choose to compute this quantity in order to test our approximation method for the simplest many-electron system, for which there are rich experimental data available \cite{weissler1955, axelrod1959, baker1961, samson1964}.
The experimental results were consolidated by Marr and West \cite{marr1976}, later by Samson, Yin and Haddad \cite{samson1994} with an accuracy of 2~\% and Bizau and Wuilleumier \cite{bizau1995}. 
We prefer to use the data by Samson \emph{et al.} \cite{samson1994} for comparison with our results due to its accuracy.

In our computations we used both the Hartree-Fock and full CI wavefunctions to describe the ground state.
The problem of the ion was trivial since the ion is a one electron system.
Both problems were solved with the help of the \textsc{GAMESS} package \cite{gamess}.
We used the basis set of Gaussian type orbitals from Ref.~\cite{przybytek2012}, called d5Z by the authors.
The number of GTO used was 86 and the ground state variational energy was at the level $E_{\mathrm{HF}} = -2.8618$~a.u. for the Hartree-Fock wavefunction and $E_{\mathrm{CI}} = -2.9033$~a.u. for the full CI wavefunction.
The configuration interaction energy is consistent with the value of $E = - 2.9037$~a.u. reported in the literature~\cite{decleva1995}.

\begin{figure}
    \centering
    \includegraphics[width=0.48\textwidth]{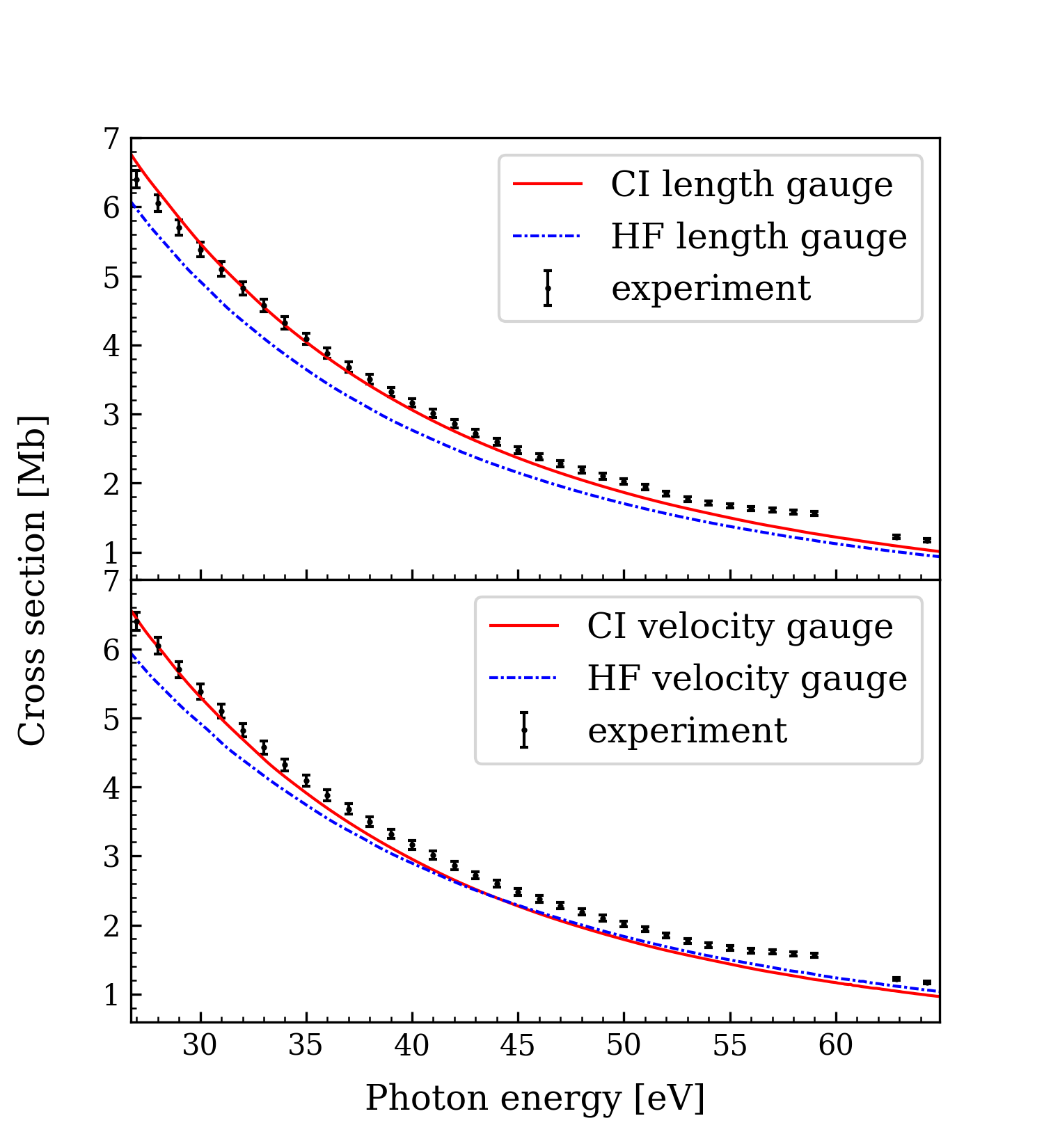}
    \caption{Photoionization cross section of the helium atom. The experimental results are taken from Samson \emph{et al.} \cite{samson1994}.}
    \label{fig:atom_he}
\end{figure}

The continuum orbital was constructed in the same way as in the case of hydrogen atom, except that now we truncated the pseudo-partial wave expansion of the Coulomb wave function (\ref{eq:approx_coulomb_wf}) at $l=6$.
The final state wavefunction was constructed according to Eq. (\ref{eq:final_state}).
The results of computation with this setup using the length and velocity form of the expression for the cross section are presented in Figure \ref{fig:atom_he}.

We achieved a good agreement with the experimental data using the length gauge form of the cross section.
The agreement is present up to 40~eV.
We also note that the results are significantly better in the case of full CI initial ground state.

\subsection{Hydrogen molecule} \label{sec:hydrogen_molecule}

The reason for the interest in the photoionization cross section of the hydrogen molecule is its presence in the interstellar gas \cite{yan1998}, as it was in the case of the helium atom.
The early computations were based on the Born-Oppenheimer approximation, where the total wavefunction was written as a product of the electronic wavefunction and the nuclear one.
Flannery and \"{O}pik used the Weinbaum's approximation for the electronic wavefunction \cite{flannery1965}.
The later calculations by Ford \emph{et al}. \cite{ford1975} extended the results to the bound and continuum vibrational levels of hydrogen ion.
All these computations were based on the following formula for the cross section
\begin{eqnarray*}
\sigma_{\nu J} =&& \frac{4 \pi^2 \omega}{3 c} \abs{ \int_0^\infty \dd{R} P_{\nu J}(R) \, \va{M}(\omega, R) \, P_{\nu_0 J_0}(R) }^2 \\
&&\times S(J, J_0),
\end{eqnarray*}
where $P_{\nu J}$ and $P_{\nu_0 J_0}$ are the rovibrational functions of nuclear motion of the final and initial states, $\va{M}(\omega, R)$ is the electronic dipole moment for the nuclei fixed at distance $R$, and $S(J, J_0)$ is an angular factor such that $\sum_J S(J, J_0) =1$.
This formula is a subject to a further simplification, e.g. the independence of $\va{M}(\omega, R)$ on $R$, called the fixed-nuclei approximation.
More elaborate calculations from vibrationally excited hydrogen molecule, including the proper average with the vibrational function were reported by the Tai and Flannery \cite{tai1977}.
The authors included the corrections to the Coulomb potential arising from the dipole moment of the ion.
Itikawa \emph{et~al.} \cite{itikawa1983} used the multi-configuration Hagstorm-Shull type wavefunction as the initial wavefunction and averaged the dipole moment operator with the vibrational wavefunctions.
Richards and Larkins \cite{richards1986} solved the Schr\"{o}dinger equation for hydrogen molecule numerically in the prolate spheroidal coordinates.

The recent calculation by Semenov and Cherepkov \cite{semenov1998} utilizes the random phase approximation with exchange \cite{schirmer1996}.
Let us note that this method became very successful as mentioned in Sec. \ref{sec:introduction}.
Another recent computation of the cross section by Morita and Yabushita \cite{morita2008} with the help of complex Gaussian basis set yielded a good agreement with the experiments.
Their method includes a way of optimizing the GTO set by expressing the functional for the frequency dependent polarizability as a function of the orbital exponents and seeking its stationary point.
Let us also note the recent discussion of Zimmermann \emph{et al.} \cite{zimmermann2015} on the effect of non-dipole photoionization of the hydrogen molecule.

The early interest in the photoelectron angular distribution was focused on the polarization averaged case~\cite{southworth1982}.
The interest in the molecular frame photoelectron distribution (MFPAD) increased, because of new experimental capabilities~\cite{ito2000, hikosaka2003}.
This stimulated theoretical investigations of this problem~\cite{semenov2003, matsuzaki2017}.

In our computations we used the same setup of the initial and final states as in the case of helium atom in Sec. \ref{sec:helium_atom}.
We choose the aug-mcc-pV5Z basis set \cite{basissetexchange}.
The variational energy of initial state at Hartree-Fock level was $E_{\mathrm{HF}} = -1.1336$~a.u. and at full CI level $E_{\mathrm{CI}} = -1.1711$~a.u., with the internuclear distance of $R= 1.4$~a.u.
The near-exact electronic energy reported in the literature for the same internuclear distance is $E = -1.1745$~a.u.~\cite{kolos1968}.
The Coulomb wavefunction was prepared in the similar fashion as in the previous case, namely we truncated the pseudo-partial wave expansion at $l=6$.

In order to test the quality of the wavefunctions of the ionic hydrogen we calculated the energies of ground and the first five exited states with the chosen basis set for the inter nuclear distance $R= 1.5$~a.u.
We compared the results with the potential curves for $\mathrm{H}_2^+$ obtained by Madsen and Peek \cite{madsen1970}.
The agreement of 3 significant digits was obtained for all states, thus we are confident that this part of the problem is described well.

Since we deal here with the molecule we need to take into account a nuclear positions.
We work in the molecular frame of reference, thus we do not consider the overall rotations of the system.
The nuclei are set to move on the $z$-axis.
We write the initial and final wavefunctions in the adiabatic approximation \cite{kolos1970}
\begin{align*}
\Psi_i (\{ \va{r} \}, R) &= P_{n_0 \nu_0} (R) \, \psi_{n_0} (\{ \va{r} \}; R), \\
\Psi_f (\{ \va{r} \}, R) &= P_{n \nu} (R) \, \psi_n (\{ \va{r} \}; R),
\end{align*}
where $R$ is internuclear separation.
The dipole moment between such states reads
$$
\int \dd{R} P_{n \nu} (R) \, P_{n_0 \nu_0} (R) \,  \va{M}_{n n_0} (R),
$$
where $\va{M}_{n n_0} (R)$ is the electronic dipole moment at a given inter nuclear separation.

Obviously we also need to sum over all possible vibrational channels.
We use the fixed nuclei approximation, namely we assume that $\va{M}_{n n_0} (R)$ is independent of $R$.
Furthermore, we assume that the sum over the final vibrational states is equal to one
$$
\sum_\nu \abs{ \int_0^\infty \dd{R}  P_{n \nu} (R) \, P_{n_0 \nu_0} (R) \, }^2 = 1.
$$
This allows to write the differential cross section to the electronic channel $n$ of the ion as 
$$
\pdv{\sigma_n}{\Omega_{\va{k}}} = \frac{4 \pi^2 \omega}{c} \, \abs{ \, \va{j} \vdot \va{M}_n(\va{k}) \, }^2, 
$$
where $\va{M}_n(\va{k})$ is electronic dipole moment, independent of $R$, and specified by the asymptotic momentum $\va{k}$ of the ejected electron and the ion electronic channel $n$. The total cross section is defined as 
$$
\sigma_{tot} = \sum_n \int \dd{\Omega_{\va{k}}} \, \,  \frac{1}{4 \pi}\int \dd{\Omega_{\va{j}}} \, \pdv{\sigma_n}{\Omega_{\va{k}}},
$$
where $n$ runs over all accessible electronic channels and the last integral averages over the polarization direction $\va{j}$.
This step is required because the total cross section is measured for an ensemble of randomly oriented molecules.

Let us note that the Coulomb wavefunction $\psi_{\va{k}}^-$ contains all the partial waves, i.e.,
it contains all necessary symmetry components, non-vanishing due to the selection rules.
This is important since one can construct the cross section as a sum over all the possible transitions (taking the selection rules into account).
Here we need not to consider such a sum since the appropriate momentum boundary conditions include the proper summation over the partial waves. 

In our case the symmetry group of the total system is $D_{\infty h}$.
This accounts for the possible angular momentum of the ejected electron in the case of $\sigma_g^+$ ionic state
\begin{align*}
m &= 0 \qquad  \; \;      l \: \mathrm{ odd} \: \: \qfor \mathrm{ parallel~transitions}, \\
m &= \pm 1 \qquad         l \: \mathrm{ odd} \: \: \qfor \mathrm{ perpendicular~transitions},
\end{align*}
and in the case of $\pi_u$ ionic state
\begin{align*}
m &= \pm 1 \quad \; \; \; l \: \mathrm{ even} \: \qfor \mathrm{ parallel~transitions}, \\
m &= 0, \pm 2 \quad       l \: \mathrm{ even} \: \qfor \mathrm{ perpendicular~transitions}.
\end{align*}

\begin{figure}
    \centering
    \includegraphics[width=0.48\textwidth]{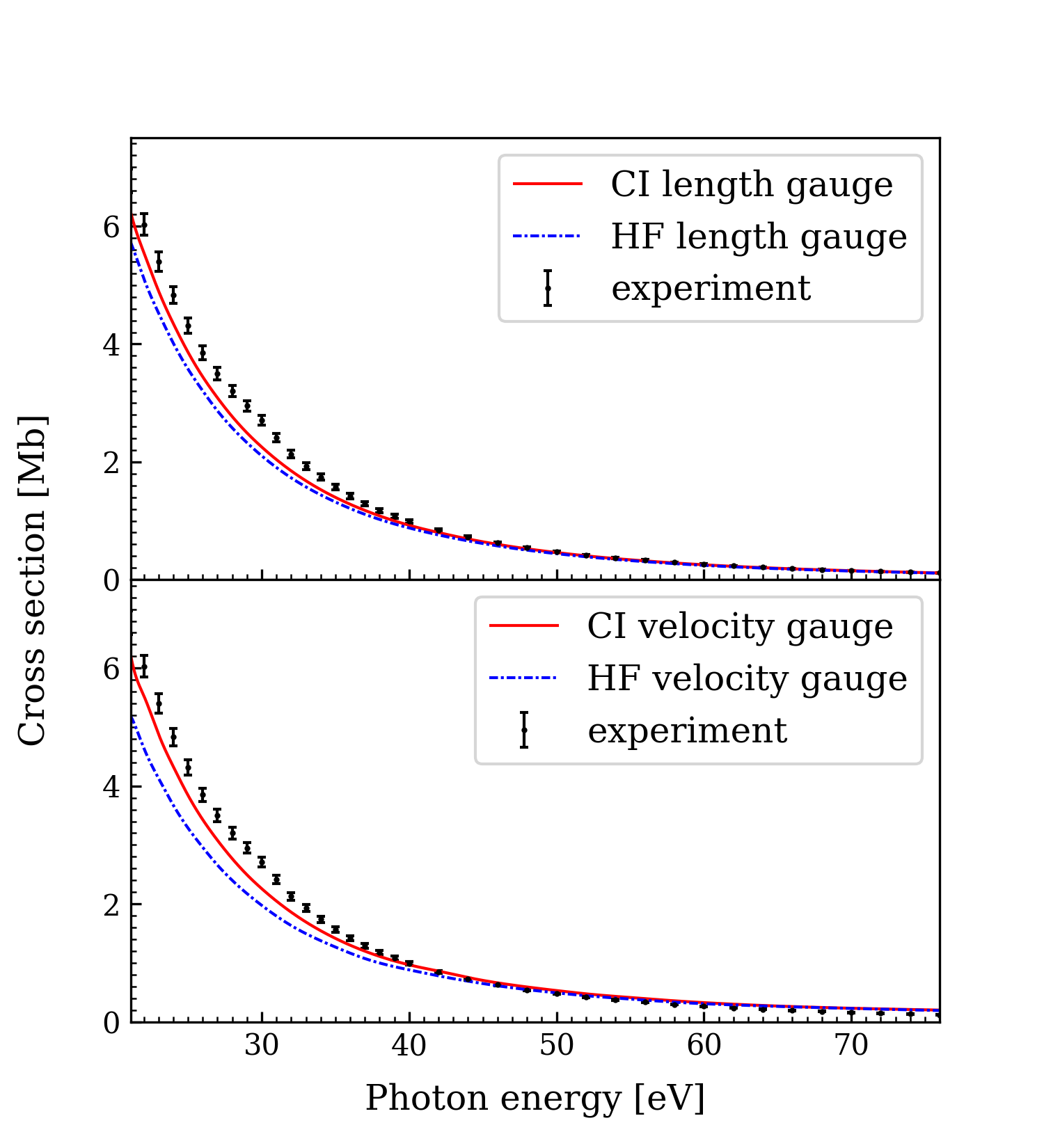}
    \caption{Total photoionization cross section of $\mathrm{H}_2$. The experimental results by Samson and Haddad \cite{samson1994_h2}.}
    \label{fig:mol_h2}
\end{figure}

In Figure \ref{fig:mol_h2}
we compare the total cross section obtained with help of our model with the data of Samson and Haddad \cite{samson1994_h2}. The results agree well with the experimental data in the region 40 - 50~eV. 
For smaller energies our method predicts smaller values of cross section, presumably due to the error contained in the ionization potential. As expected, the results are substantially more accurate when the full CI expansion is used to represent the ground-state wavefunction.
In the case of Hartree-Fock ground state the length gauge computation yields more reliable cross section.

In our computation we aimed at reproducing the photoelectron angular distribution.
We choose the low energy experimental data by Hikosaka and Eland \cite{hikosaka2003} as a reference where only one electronic channel is open, $1s \sigma_g^+$ of $\mathrm{H}_2^+$.
We also decided to compare our results with the measurements of Ito \emph{et al.} \cite{ito2000}, of PADs into the excited $2p \pi_u$ and $2s \sigma_g^+$ states of $\mathrm{H}_2^+$ for photon energies 44 -76~eV.

\begin{figure*}
\begin{subfigure}{0.49\textwidth}
    \centering
    \includegraphics[width=\textwidth]{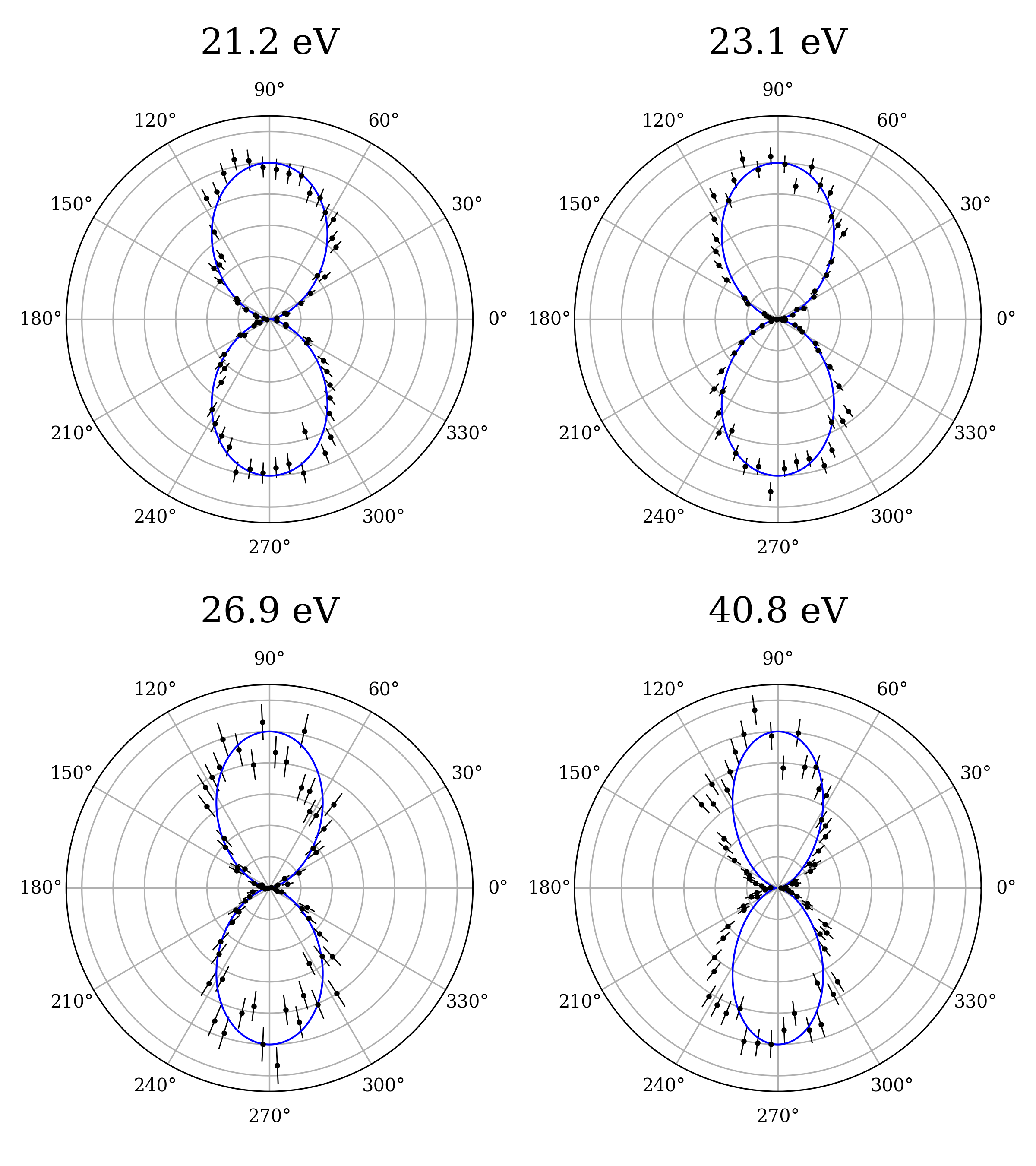}
    \caption{The computation performed within the velocity gauge.}
    \label{fig:perpendicular_vel}
\end{subfigure}
\begin{subfigure}{0.49\textwidth}
    \centering
    \includegraphics[width=\textwidth]{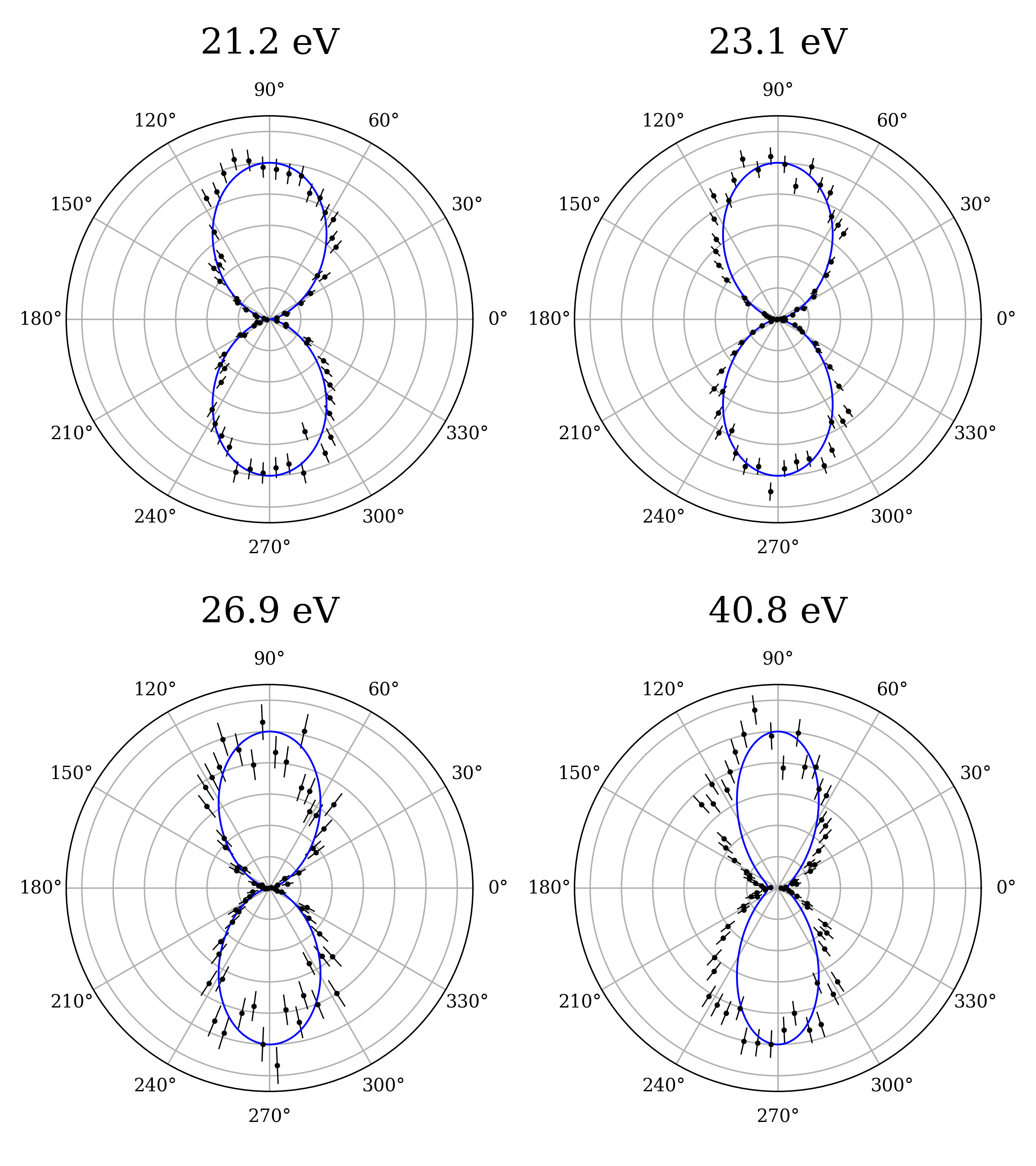}
    \caption{The computation performed within the length gauge.}
    \label{fig:perpendicular_dip}
\end{subfigure}
    \caption{The MFPAD of $\mathrm{H}_2$. The molecule lies on the axis $0^\circ$-$180^\circ$ and the polarization vector lies on the line $90^\circ$-$270^\circ$. The results are represented by blue lines and were obtained as interpolation of the computed cross section. The experimental results by Hikosaka and Eland \cite{hikosaka2003}, marked as black errorbars.}
\end{figure*}

\begin{figure*}
\begin{subfigure}{0.49\textwidth}
    \centering
    \includegraphics[width=\textwidth]{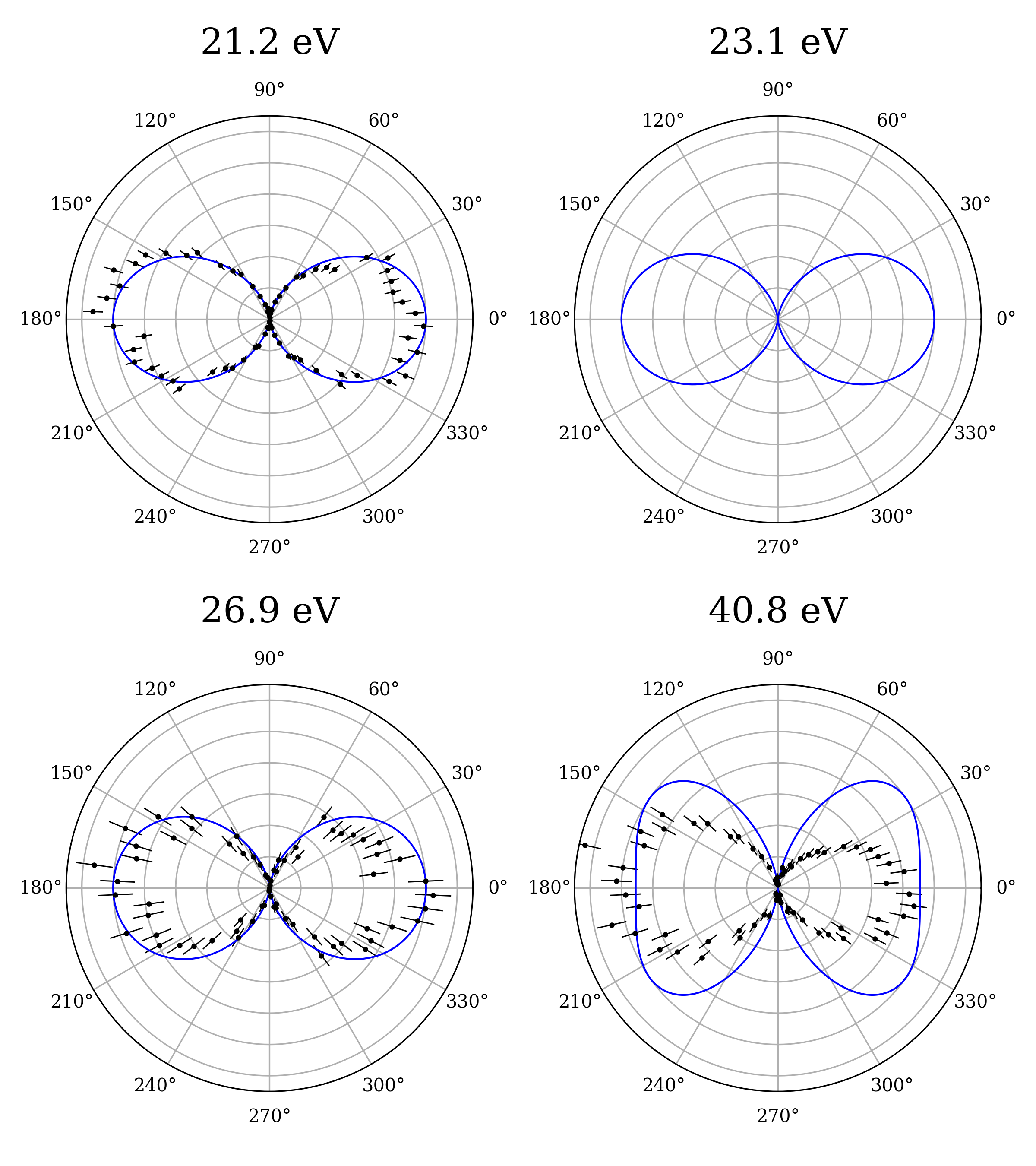}
    \caption{The computation performed within the velocity gauge.}
    \label{fig:parallel_vel}
\end{subfigure}
\begin{subfigure}{0.49\textwidth}
    \centering
    \includegraphics[width=\textwidth]{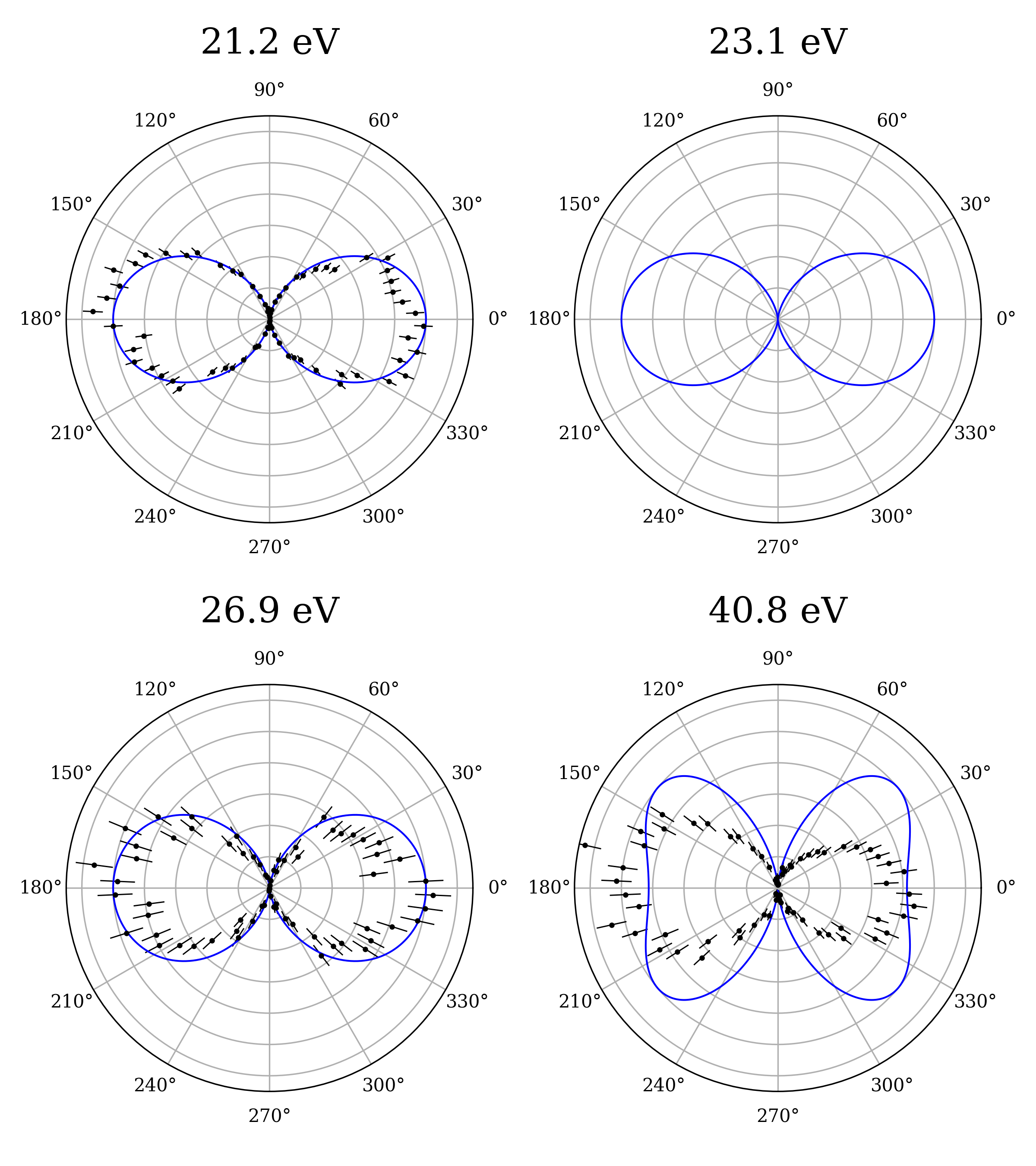}
    \caption{The computation performed within the length gauge.}
    \label{fig:parallel_dip}
\end{subfigure}
    \caption{The MFPAD of $\mathrm{H}_2$. The molecule lies on the axis $0^\circ$-$180^\circ$ and the polarization vector lies on the line $0^\circ$-$180^\circ$. The results are represented by blue lines and were obtained as interpolation of the computed cross section. The experimental results by Hikosaka and Eland \cite{hikosaka2003}, marked as black errorbars.}
\end{figure*}

\begin{figure*}
\begin{subfigure}{0.49\textwidth}
    \centering
    \includegraphics[width=\textwidth]{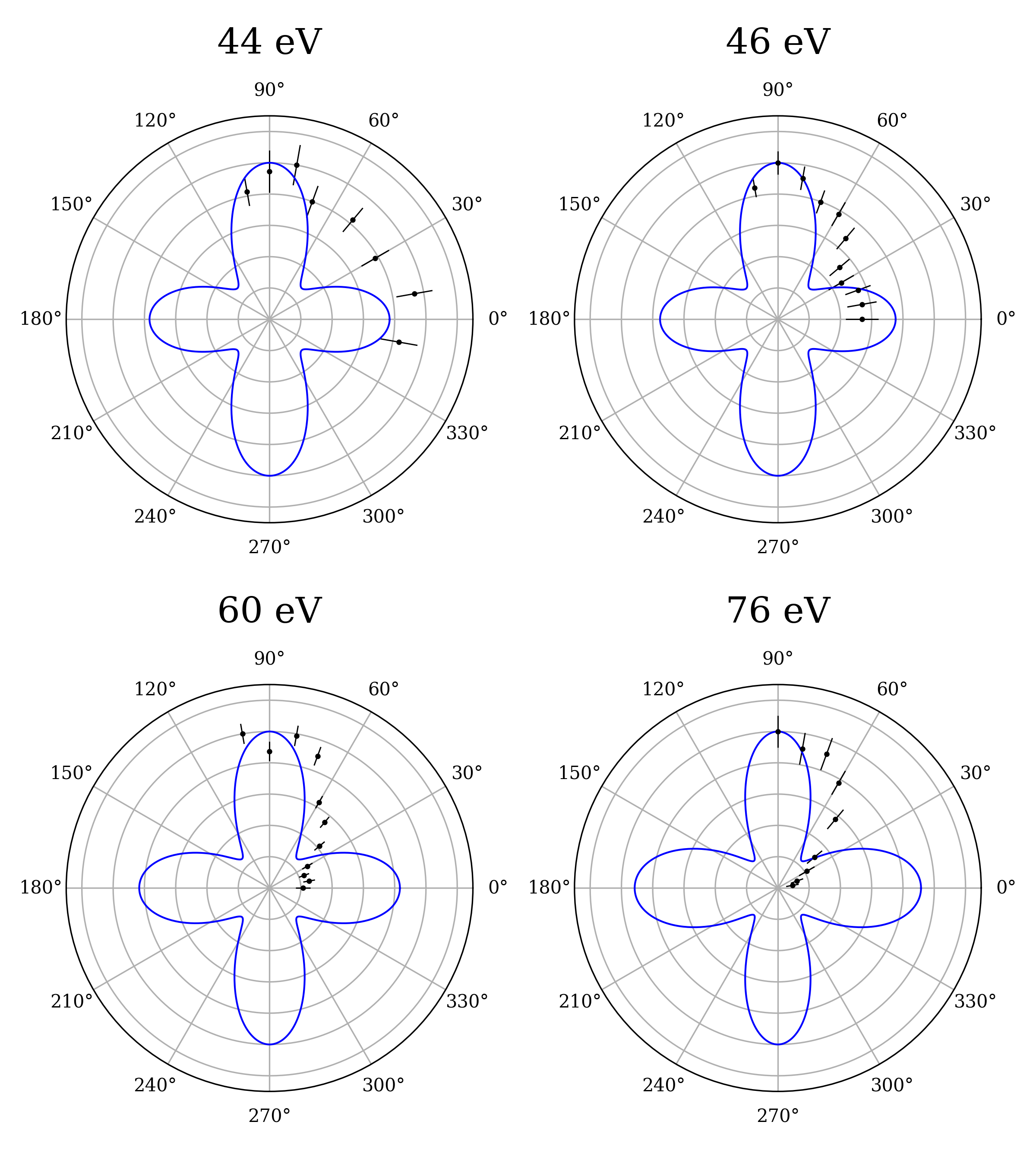}
    \caption{The computation performed within the velocity gauge.}
    \label{fig:parallel_vel_ito}
\end{subfigure}
\begin{subfigure}{0.49\textwidth}
    \centering
    \includegraphics[width=\textwidth]{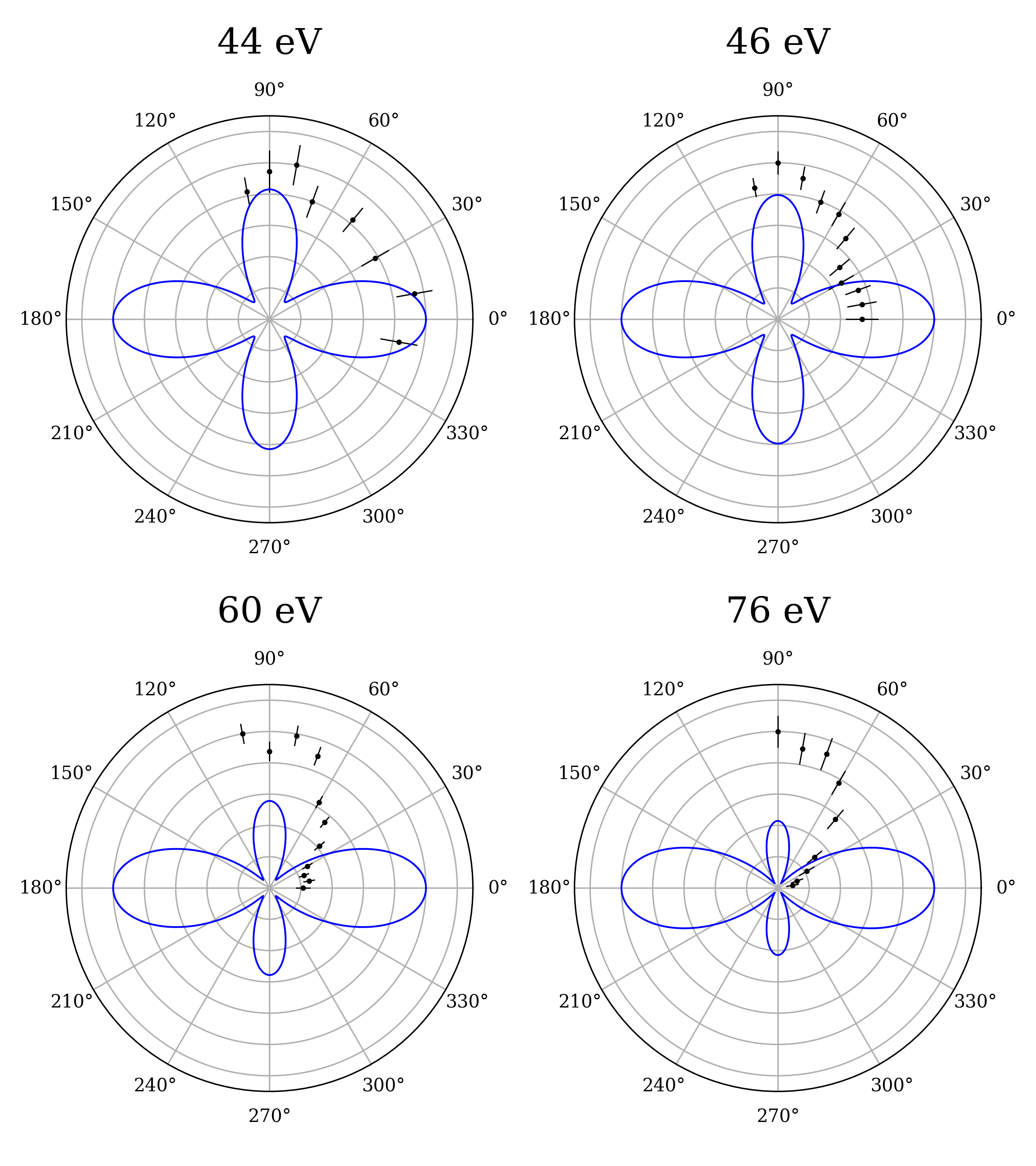}
    \caption{The computation performed within the length gauge.}
    \label{fig:parallel_dip_ito}
\end{subfigure}
    \caption{The MFPAD of $\mathrm{H}_2$ into the $2p \pi_u$ and $2s \sigma_g^+$ of the ion states. The molecule lies on the axis $0^\circ$-$180^\circ$ and the polarization vector lies on the line $90^\circ$-$270^\circ$. The results are represented by blue lines and were obtained as interpolation of the computed cross section. The experimental results by Ito \emph{et al.}~\cite{ito2000}, marked as black errorbars.}
\end{figure*}

We present the results of the photoionization computations for the ionic $1s \sigma^+_g$ state in the case of polarization vector $\va{j}$ oriented perpendicular to the molecular axis in Figures \ref{fig:perpendicular_vel} and \ref{fig:perpendicular_dip}.
The results for the case of the parallel alignment are shown in Figures \ref{fig:parallel_vel} and~\ref{fig:parallel_dip}.
The comparison to the data for excited $2p \pi_u$ and $2s \sigma_g^+$ ionic states are presented in Figures \ref{fig:parallel_vel_ito} and \ref{fig:parallel_dip_ito}.
The authors of Ref.~\cite{ito2000} only published the data in the case of perpendicular polarization with respect to the molecular axis, thus we do not show the results for the parallel alignment.


The computations were done with the initial state of hydrogen molecule taken as a full CI wavefunction using both the length and velocity gauges of the transition moment.
We computed the cross section for photon energies 21.2, 23.1, 26.9, 40.8, 44.0, 46.0, 60.0 and 76.0~eV in accordance with the energies of the experimental results.
However, we have no access to the results at the photon energy 23.1~eV in the parallel case, so the experimental results are absent for this energy in Figures~\ref{fig:parallel_vel} and~\ref{fig:parallel_dip}.

Let us assume that the molecular axis coincides with the $z$-axis of the coordinate system.
Since the hydrogen molecule has the cylindrical symmetry we consider only the azimuthal angle $\theta$ of the asymptotic momentum $\va{k}$ of the ejected electron.
Moreover, there is also a symmetry plane perpendicular to the $z$-axis, thus it is sufficient to consider the azimuthal angle $\theta$ of $\va{k}$ in the domain $ [ 0, \pi /2 ] $.
We computed the cross section for ten evenly spaced values of $\theta$ in $ [ 0, \pi /2] $.
In the plots the interpolating curves over these values are shown, reflected in the $0^\circ$-$180^\circ$ axis by the cylindrical symmetry and in the $90^\circ$-$270^\circ$ axis by the symmetry plane.
Since the experimental data are given as relative intensities, the plots were normalized to the unity in the maximum.
The experimental data of Ito \emph{et al.} \cite{ito2000} were normalized in a similar fashion.
The data of Hikosaka and Eland \cite{hikosaka2003} were normalized such that the average of four maximal values is equal to the unity.

Our results agree well with the experimental data of Hikosaka and Eland \cite{hikosaka2003}, with the exception of  energy 40.8~eV in the parallel case in Figures~\ref{fig:parallel_vel} and~\ref{fig:parallel_dip}, where our results exhibit d wave shape contribution.
However, this feature is consistent with the previous theoretical computations of Semenov and Cherepkov~ \cite{semenov2003}.

\section{Conclusions \label{sec:conclusions}}
In this paper we developed a new method of describing stationary continuum states with the help of a discrete basis set.
Our method is based on the assumption that the ejected electron can by described by Coulomb wavefunction.
We present the procedure of generating the approximated continuum orbitals, suitable for obtaining the photoelectron angular distribution.
Moreover, our method strives to be compatible with the existing quantum chemistry software, by employing the use of the Gaussian type orbitals.
We also compare the results obtained with our method against experiments as well as other computational techniques.
We believe that our approach can be extended to more sophisticated frameworks of obtaining the photoelectron angular distribution as well as time-dependent problems.

\section*{Acknowledgments}
We acknowledge the Polish National Science Center Grant No. 2016/20/W/ST4/00314.

\bibliography{main}
\end{document}